**Conformal Slit Mapping Based Spiral Tool Trajectory Planning for Ball-end Milling on Complex Freeform Surfaces**

Changqing Shen [a], BingZhou Xu [a], Xiaojian Zhang [a]*, Sijie Yan [a], Han Ding [a]

[a]State Key Laboratory of Intelligent Manufacturing Equipment and Technology, School of Mechanical Science and Engineering, Huazhong University of Science and Technology, Wuhan, 430074, China

***Corresponding author**

Xiaojian Zhang, State Key Laboratory of Intelligent Manufacturing Equipment and Technology, School of Mechanical Science and Engineering, Huazhong University of Science and Technology, No. 1307, Luoyu Street, Hongshan District, Wuhan, 430074, China

E-mail: xjzhang@hust.edu.cn; Phone 13554692064.

**Abstract**

This study presents a spiral-based complete coverage strategy for ball-end milling on freeform surfaces, utilizing conformal slit mapping to generate milling trajectories that are more compact, smoother, and evenly distributed when machining 2D cavities with islands. This approach, an upgrade from traditional methods, extends the original algorithm to effectively address 3D perforated surface milling. Unlike conventional algorithms, the method embeds a continuous spiral trajectory within perforated surfaces without requiring cellular decomposition or additional boundaries. The proposed method addresses three primary challenges, including modifying conformal slit mapping for mesh surfaces, maintaining uniform scallop height between adjacent spiral trajectories, and optimizing the mapped origin point to ensure uniform scallop height distribution.

To overcome these challenges, surface flattening techniques are incorporated into the original approach to accommodate mesh surfaces effectively. Tool path spacing is then optimized using a binary search strategy to regulate scallop height. A functional energy metric associated with scallop height uniformity is introduced for rapid evaluation of points mapped to the origin, with the minimum functional energy determined through perturbation techniques. The optimal placement of this point is identified using a modified gradient descent approach applied to the energy function. Validation on intricate surfaces, including low-quality and high-genus meshes, verifies the robustness of the algorithm. Surface milling experiments comparing this method with conventional techniques indicate a 15.63% improvement in scallop height uniformity while reducing machining time, average spindle impact, and spindle impact variance by up to 7.36%, 27.79%, and 55.98%, respectively.

## 1. Introduction

### 1.1. Problem Definition

Efficient trajectory planning for ball-end milling plays a vital role in manufacturing intricate surfaces in molds, automotive components, and aerospace structures [1–3]. Various trajectory strategies have been explored [4–6], with spiral trajectories gaining widespread adoption [7–9], due to their distinct advantages. These include smooth direction transitions, low interruptions and intersections, and effective boundary alignment, all of which contribute to shorter milling durations, fewer machining defects, and low residual material along boundaries.

Based on the principles of conformal slit mapping [10–13], the conformal slit mapping-based spiral complete coverage path planning (CSM-SCCPP) approach was introduced [14]. This method facilitates the creation of a continuous spiral trajectory within a 2D multiply connected region without necessitating additional boundary constraints for subregion decomposition. Consequently, it generates more compact trajectories, smoother turns, and improved spacing uniformity compared to conventional techniques that depend on subregion decomposition for handling multiple connected domains [15,16]. Extending CSM-SCCPP to ball-end milling for machining complex surfaces holds promise for improving both quality and efficiency. However, its application to such scenarios introduces three key challenges:

Limitations in surface representation: Complex surfaces are commonly described using triangular meshes, while the CSM-SCCPP method depends on the generalized Neumann kernel approach [11,12] to compute conformal slit mappings. However, this approach is designed to handle only 2D boundary inputs and cannot directly process triangular mesh surfaces.

Optimization of trajectory spacing: Inadequate trajectory spacing can result in either excessive scallop height or unnecessarily prolonged toolpaths. The spacing produced by CSM-SCCPP may experience sudden variations due to surface discontinuities, making conventional scallop height equation-based spacing control methods ineffective. Accordingly, identifying the optimal trajectory spacing for machining complex surfaces remains a challenging and active research area.

Placement of the origin-mapped point: In conformal slit mapping, selecting a reference point that is mapped to the origin plays a crucial role in determining the uniformity of scallop heights along the milling trajectory. However, an optimal strategy for positioning this point remains unexplored.

To overcome these challenges, this study expands CSM-SCCPP for ball-end milling of complex surfaces by modifying conformal slit mapping with the generalized Neumann kernel method to accommodate triangular mesh surfaces. Trajectory spacing is optimized according to predefined scallop height constraints, while an effective strategy is developed for selecting the origin-mapped point to achieve uniform scallop heights within a reasonable computational time frame of minutes.

### *1.2. Highlight*

To facilitate the transition from 2D plane milling to complex surface ball-end milling, CSM-SCCPP is enhanced with a crucial innovation: an efficient method for determining the optimal placement of the origin-mapped point in conformal slit mapping. This optimization establishes a minimal functional energy criterion and utilizing gradient descent to locate the ideal position along the energy landscape. This improvement enables the generation of uniformly spaced trajectories with consistent scallop heights with processing times within an acceptable range of minutes.

### *1.3. Related work*

In practical manufacturing, triangular surface meshes are typically obtained from surface models or reconstructed from scanned point clouds and can be flattened (parameterized) using various techniques [17,18]. By first planning milling tool trajectories on these flattened representations and subsequently mapping them back onto the original freeform surfaces, the complexity of handling trajectory intersections and discontinuities is significantly reduced, effectively converting the problem into a simpler 2D task [19,20]. This approach facilitates the design of smooth, well-structured, and complete coverage trajectories, making it a widely adopted and effective strategy for generating milling paths on freeform surfaces.

Various surface flattening algorithms preserve angles [21,22] and areas [23] to varying extents, but aside from a few developable surfaces, achieving perfect isometric flattening by maintaining properties simultaneously is not feasible [24]. In CNC machining, where trajectory smoothness is critical [25], ensuring that smoothly planned 2D tool paths remain smooth when remapped onto freeform surfaces has led researchers to favor conformal flattening methods that preserve angles [7,26,27]. One such method, conformal slit mapping (CSM), transforms mesh surfaces with holes into structured annular regions with arc slits by computing holomorphic 1-form bases [10,13], facilitating the generation of spiral trajectories within these mapped regions. CSM allows discrete control over the central mapped position by removing a triangular facet from the surface. However, if the removed facet is excessively elongated, severe distortion may occur in the mapping near the center. Alternatively, a 2D CSM approach based on the generalized Neumann kernel provides continuous adjustments to the central mapped position without computational distortions, though it lacks direct support for mesh-based inputs [11,28]. Other conformal flattening techniques, such as boundary-free approaches [21] like the boundary first flattening (BFF) algorithm [22] and boundary-fixed methods such as annular conformal mapping [27,29], also exist. However, these methods confront challenges in mapping internal holes into easily manageable regular regions, limiting their effectiveness primarily to trajectory planning on surfaces without holes.

Once complex surfaces are flattened, trajectory planning simplifies into a 2D problem. Spiral trajectories are widely used in 2D cavity machining due to their smooth transitions, low discontinuities, absence of self-intersections, and adaptability to intricate boundaries. This approach can reduce machining impact, shorten tool lift times, and reduce scallops or surface scratches. Research indicates that conformally mapping spiral trajectories from 2D to 3D can significantly improve milling quality and efficiency on freeform surfaces [27]. However, conventional methods face challenges in achieving full coverage in multiple connected regions, often resulting in numerous discontinuities, abrupt turns, and uneven trajectory spacing [16,30]. In contrast, the CSM-SCCPP method utilizes the iso-parametric lines of CSM to achieve coverage of multiple connected regions while effectively avoiding holes. This method ensures smooth trajectory transitions and maintains a concentrically arranged structure that facilitates bridging between different trajectory segments, leading to high-quality coverage paths. Despite these advantages, its application has not yet been extended to 3D surfaces.

During the machining of surfaces with ball-end mills, the height of milling scallops is affected by the spacing between tool paths, making it essential to impose constraints on trajectory spacing to regulate scallop height. Two principal methods are utilized to determine this spacing. The first method calculates scallop height and trajectory spacing based on the ball-end mill's radius and the local curvature of the surface. Research in this area includes the development of a Riemannian metric for trajectory spacing, formulated using scallop height equations [31]. Accordingly, Zou introduced an optimization strategy that ensures the scalar field gradients remain nonzero on surfaces without holes, thereby achieving globally uniform scallop heights and smoothly transitioning tool paths. However, applying this technique to surfaces with holes remains challenging, as constructing scalar fields with nonvanishing gradients in such regions is an unresolved issue [32]. Additionally, Lee observed that equal-scallop trajectories generated incrementally often exhibit excessive interruptions and abrupt direction changes [33]. Furthermore, the approach does not inherently prevent tool path intersections with holes on the surface. The second method determines trajectory spacing through machining simulation. This technique involves deriving the machined surface profile by subtracting the area swept by the tool from the rough workpiece [34]. In cases where an in-depth analysis of scallop shapes is unnecessary, the rough blank can be approximated using iso-scallop surfaces, a simplification that reduces computational complexity [35]. To achieve optimal spacing, it is crucial to ensure that the entire iso-scallop surface remains confined within the tool's swept area, thereby maintaining the intended maximum scallop height. The spacing can be refined through an iterative bisection method [34]. However, precise optimization requires discretizing the iso-scallop surfaces into a densely distributed set of points [36]. Despite the use of computational acceleration techniques such as KD-trees, which expedite the calculation of minimum distances between point sets [37].

## 2. Method

### *2.1. Conformal slit mapping for mesh surfaces*

This section introduces a technique for transforming mesh surfaces into unit disks or unit annular domains with slits on the complex plane. The approach utilizes surface conformal flattening methods [22] incorporated with a 2D conformal slit mapping algorithm based on the generalized Neumann kernel method [11,28]. Fig. 1 illustrates the implementation of the algorithm.

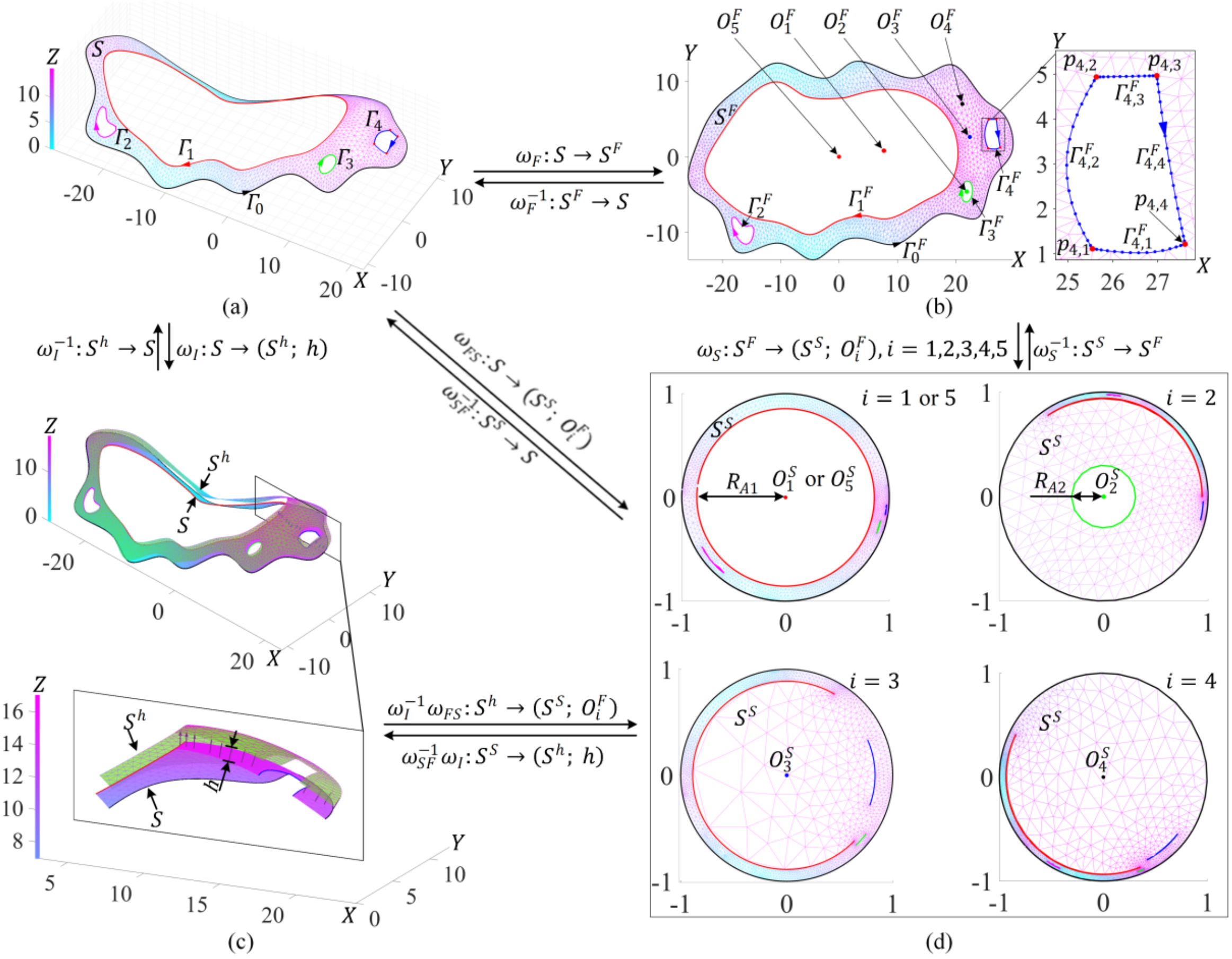

**Fig. 1.** Transformation relationships between different mappings: **(a)** The original surface $S$, **(b)** The flattened surface $S^F$ obtained from the surface $S$, **(c)** The iso-scallop surface $S^h$ obtained by offsetting the surface $S$ by a distance $h$ normal to the surface, **(d)** The surface $S^F$ mapped to a disk or annular region $S^S$ using conformal slit mapping, with $O^F$ positioned at different locations.

The mesh surface $S$ consists of $m + 1$ ($m > 0$) boundaries, denoted as $\Gamma_i$, ($i = 0,1,\ldots,m$). Among these, a specific boundary, $\Gamma_0$, is selected for mapping via conformal slit mapping to form the outer boundary of either the unit disk domain $\overline{D}$ or the unit annular domain $\bar{A}$. Typically, the longest boundary is chosen as $\Gamma_0$.

Initially, the BFF algorithm [22] is employed to generate a flattened representation of the surface, denoted as $S^F$. The mapping function for this flattening process is represented as $\omega_F: S \to S^F$. The boundaries of $S^F$ corresponding to the original surface are labeled as $\Gamma_i^F$, $i =$ $0,1,\ldots,m$. As illustrated in Fig. 1(b), $\Gamma_i^F$ may exhibit either smooth continuity ($C1$-continuous) or a piecewise smooth structure, incorporating a finite number of $C1$-discontinuous corner points.

Based on previous research on 2D conformal slit mapping [11,14,38], the boundary $\Gamma^F = \Gamma_0^F \cup \Gamma_1^F \cup \ldots \cup \Gamma_m^F$ is parameterized using a 2π-periodic parameter $t$. This parameterization is defined as $\Gamma_i^F(t)$, where $t \in J_i = [0, 2\pi]$ and $i = 0,1,\ldots,m$. The formulation adheres to the standard approach outlined in Equation A-14 of Shen et al. [14].

When $\Gamma_i^F(t)$ includes a finite number of $C1$-discontinuous corner points, the Nyström parameterization method [38] is employed to enhance the accuracy of convergence in subsequent calculations. This approach ensures a smoother variation in arc length near corners, thereby improving the numerical stability of the conformal slit mapping solution. As depicted in Fig. 1(b), 128 uniformly distributed parameter points along $\Gamma_4^F(t)$ are defined as $\Gamma_4^F\left(\frac{i}{128} * 2\pi\right)$ for $i = 1,2,\ldots,128$. Upon completing the boundary parameterization, the conformal slit mapping of $S^F$ is constructed using two distinct methodologies based on the generalized Neumann kernel method [11].

Disk conformal mapping: A reference point $O^F$ is selected within $S^F - \Gamma^F$.

Annular conformal mapping: A reference point $Z_1$ is chosen from the region $S^F(\Gamma_1^F) + S^F(\Gamma_2^F) + \cdots + S^F(\Gamma_m^F) - \Gamma^F$, where $S^F(\Gamma_i^F)$ represents the simply connected domain enclosed by $\Gamma_i^F$.

To maintain consistency, notation is standardized such that $O^F$ is selected within $S^F(\Gamma_0^F) - \Gamma^F$. The decision to apply either disk or annular conformal slit mapping depends on the placement of $O^F$. As depicted in Figs. 1(b) and 1(d), positioning $O^F$ at $O_3^F$ or $O_4^F$ results in mapping $S^F$ onto $\overline{D}$, whereas placing $O^F$ at $O_1^F$, $O_2^F$, or $O_5^F$ maps $S^F$ onto $\bar{A}$. The conformal slit mapping of $S^F$ is denoted as $\omega_S: S^F \to (S^S;\ O^F)$.

The inverse mapping, $\omega_S^{-1}: S^S \to S^F$, is determined by converting between Cartesian and barycentric coordinates. For any point $P \in S^S$, the triangle index within $S^S$ is first identified, and $P$ is expressed in barycentric coordinates relative to that triangle. The corresponding Cartesian coordinates in $S^F$ are then retrieved by referencing the same triangle index in $S^F$, yielding $\omega_S^{-1}(P)$. Similarly, the inverse transformation $\omega_F^{-1}: S^F \to S$ is computed using the same approach.

Once $O^F$ is defined, the surface $S$ can be transformed into $S^S$, where the origin of $S^S$ is designated as $O^S$. The domain $S^S$ corresponds to either a unit disk $\overline{D}$ or a unit annular domain $\bar{A}$, with an inner circular boundary of radius $R_A$. This transformation is accomplished through the composite mapping $\omega_{FS} = \omega_F \circ \omega_S: S \to (S^S;\ O^F)$. The reverse transformation, which maps $S^S$ back to $S$, is given by the inverse composite mapping $\omega_{SF}^{-1} = \omega_S^{-1} \circ \omega_F^{-1}: S^S \to S$.

Next, consider the transformation of surface $S$ along its normal direction by a distance $I$, represented as $\omega_I: S \to (S^I;\ I)$. When $I$ corresponds to the maximum scallop height $h$, the resulting equidistant surface is denoted as $S^h = \omega_I(S, h)$, forming the iso-scallop surface. Utilizing the previously defined mappings, the transformation that maps $S^I$ to $S^S$ is expressed as $\omega_I^{-1}\omega_{FS}: S^I \to (S^S;\ O^F)$. Similarly, the inverse transformation, which maps $S^S$ back to $S^I$, is given by $\omega_{SF}^{-1}\omega_I: S^S \to (S^S;\ I)$.

*2.2. Spiral path generation through conformal slit mapping*

Let $C_S = \{C_1^S, C_2^S, \dots, C_k^S\}$ represent the set of all concentric circles on $S^S$, each centered at $O^S$, with corresponding decreasing radii $R_S = \{R_1^S, R_2^S, \dots, R_k^S\}$, as illustrated in Fig. 2(a). The set of iso-parameter curves on $S$ is denoted as $TC = \omega_{SF}^{-1}(C_S) = \{C_1, C_2, \dots, C_k\}$, where each curve is obtained through the inverse mapping $C_i = \omega_{SF}^{-1}(C_i^S)$ for $i = 1,2, \dots, k$. As shown in Figs. 2(b) and 2(d), the iso-parameter curves in $TC$ conform to the boundaries of $S$, avoid intersecting holes, and do not overlap. These curves maintain a concentric arrangement that facilitates bridging while exhibiting smooth curvature, making them well-suited as contact points for machining tools.

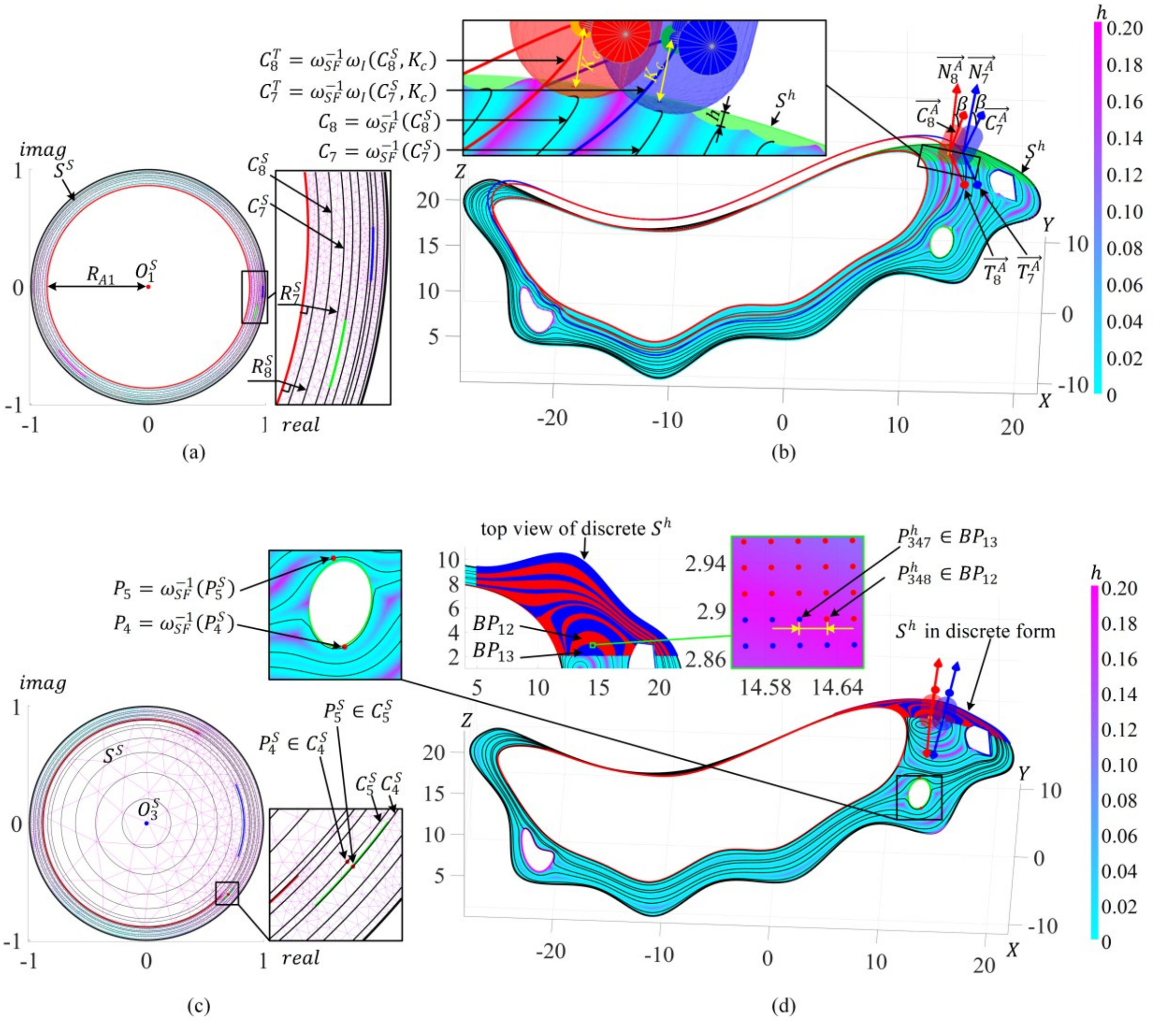

**Fig. 2.** Tool trajectory generation based on conformal slit mapping and corresponding machining scallop height. (a) and (c) Conformal slit mapping domain $S^S$ and iso-parametric curves on the mapped domain $S^S$. (b) and (d) Tool center trajectories $C_i^T$, tool contact trajectories$C_i$, tool axis direction $\overrightarrow{C_i^A}$, and the corresponding milling scallop height distribution and milling bands $BP_i$.

Given that the optimal placement of $O^F$ is predetermined, this section explores two fundamental aspects. The first involves regulating the gap between successive iso-parameter curves within $TC$ to achieve a desired maximum scallop height on the machined surface. The second focuses on strategies for linking these trajectories, with an emphasis on utilizing a spiral configuration.

2.2.1. Iso-parametric trajectories spacing control

The initial tool center trajectory, denoted as $C_1^T$, for the ball end mill is determined using the expression $C_1^T = \omega_{SF}^{-1}\omega_I(C_1^S(R_1^S), K_c)$, where $K_c$ represents the tool radius, and $R_1^S$ is a trajectory parameter selected within the interval $R_1^S \in [R_{min}, 1]$. In the case of disk conformal mappings, $R_{min} = 0$, whereas for annular conformal mappings, $R_{min} = R_A$.

Define the $P^h = \{P_1^h, P_2^h, \dots, P_{N_S}^h\}$, where each point is associated with its respective mesh elements and barycentric coordinates. These mappings are stored to eliminate redundant coordinate transformations across mesh elements. The subset $P^{h,C_1^T,K_c^+} \subset P^h$ is defined as:

$$P^{h,C_1^T,K_c^+} = \left\{P_i^h \in P^h \mid \left\|P_i^h - C_1^T\right\|_2 > K_c\right\} \tag{1}$$

This subset consists of points on $S^h$ that are positioned at a Euclidean distance exceeding $K_c$ from $C_1^T$. Utilizing the stored barycentric coordinates within the triangular mesh structure of $P^h$, the optimal value of $R_1^S$ is determined efficiently through the following procedure.

Let $P^h$ and $P^{h,C_1^T,K_c^+}$ be defined under the mapping $\omega_I^{-1}\omega_{FS}$ on $S^S$ as follows:

$$P^S = \omega_I^{-1}\omega_{FS}(P^h, O^F) \tag{2}$$

$$P^{S,C_1^T,K_c^+} = \omega_I^{-1}\omega_{FS}\left(P^{h,C_1^T,K_c^+}, O^F\right) \tag{3}$$

Next, we define the set of points $P^{P^S,C_1^S}$ that are part of $P^{S,C_1^T,K_c^+}$ but are not enclosed by the circle $C_1^S$. More specifically:

$$P^{P^S,C_1^S} = \left\{P_i^S \in P^{S,C_1^T,K_c^+} \mid \left\|P_i^S - O^S\right\|_2 > R_1^S\right\} \tag{4}$$

A binary search is then performed within the interval $[R_{min}, 1]$ to adjust the radius $R_1^S$ associated with $C_1^S$, ensuring that $P^{P^S,C_1^S}$ is exactly empty.

Once $R_1^S$ is determined, consider $P^{h,C_1^T,K_c^+}$ as $P^h$ and repeat the binary search process to find $R_2^S$. For this iteration, adjust the search range to $[R_{min}, R_1^S]$. This procedure is repeated for each successive $R_i^S$ until $P^h = \emptyset$, at which point the process concludes.

The determination of $R_i^S$ results in the calculation of the tool contact trajectory $C_i = \omega_{SF}^{-1}\left(C_i^S\left(R_i^S\right)\right)$ and the corresponding tool center trajectory $C_i^T = \omega_{SF}^{-1}\omega_I\left(C_i^S\left(R_i^S\right), K_c\right)$. At this stage, the tool axis direction $\overrightarrow{C_i^A}$ has two degrees of freedom (Sun & Altintas, 2016). In this study, a commonly used tilt angle of 0° and a lead angle of 15° are chosen to uniquely define $\overrightarrow{C_i^A}$, which can be expressed as:

$$\overrightarrow{C_i^A} = \overrightarrow{N_i^A}cos(\beta) + \overrightarrow{T_i^A}sin(\beta) \tag{5}$$

The surface normal vector at the contact point is denoted as $\overrightarrow{N_i^A}$, and the feed direction at the contact point is represented as $\overrightarrow{T_i^A}$. The tool axis lead angle is symbolized by $\beta$, as shown in Fig. 2(b).

The tool contact trajectories, tool center trajectories, and tool axis direction trajectories corresponding to $C_S$ are defined as follows:

$$\begin{cases} TC = \{\omega_{SF}^{-1}(C_1^S), \omega_{SF}^{-1}(C_2^S), \dots, \omega_{SF}^{-1}(C_k^S)\} \\ TB = \{C_1^T, C_2^T, \dots, C_k^T\} \\ TA = \left\{\overrightarrow{C_1^A}, \overrightarrow{C_2^A}, \dots, \overrightarrow{C_k^A}\right\} \end{cases} \tag{6}$$

To distinguish the milling bands of various tool center trajectories $TB$ on $S^h$, alternating red and blue colors are applied. The set of discrete points within these milling bands is represented as $BP = \{BP_1, BP_2, \dots, BP_k\}$. As illustrated in Fig. 2(d), the milling bands connect smoothly to form $S^h$, ensuring the entire surface's scallop height remains below $h$, which demonstrates the efficiency of the trajectory spacing optimization technique.

Appendix A contains the pseudocode for trajectory spacing control and an analysis of its algorithmic complexity, as this component constitutes the most computationally demanding part of the method. The complexity evaluation indicates that the trajectory spacing control algorithm operates with a time complexity of $O\left(kN_S log(N_C) log\left(\frac{1}{\varepsilon}\right)\right)$, where $k$ represents the total number of computed iso-parametric trajectories, $N_S$ corresponds to the number of sampled points $P^h$ on $S^h$, $N_C$ signifies the count of discrete points on $C_i^S$, and $\varepsilon$ denotes the iteration tolerance governing the spacing between successive trajectories.

2.2.2. Iso-parametric trajectories spiral bridging

Since the tool contact trajectories $TC$ are arranged coaxially, they can be seamlessly connected without abrupt intersections or sharp directional changes. This configuration minimizes unnecessary tool retractions and re-engagements, effectively reducing machining time and surface imperfections. The following approach is employed to achieve this:

The domain $S^S$ is transformed into a rectangular region $S_0^R$ using the mapping defined as follows:

$$S_0^R = arg(S^S) + i|S^S| \tag{7}$$

Let the mapping be represented as $\omega_R: S^S \rightarrow S_0^R$. Under this transformation, the arc-shaped slits present in $S^S$ are converted into linear slits within $S_0^R$. By shifting and replicating $S_0^R$ along the real axis with a periodic interval of 2π, a series of regions denoted as $S_i^R$ is generated. The subscript $i = -1,0,1, \ldots$ indicates that each region $S_i^R$ is obtained by translating $S_0^R$ by a displacement of $2\pi i$ along the real axis. The union of these translated regions $S^R$, represented as $S_{-1}^R \cup S_0^R \cup S_1^R \cup \ldots$, is illustrated in Fig. 3(a).

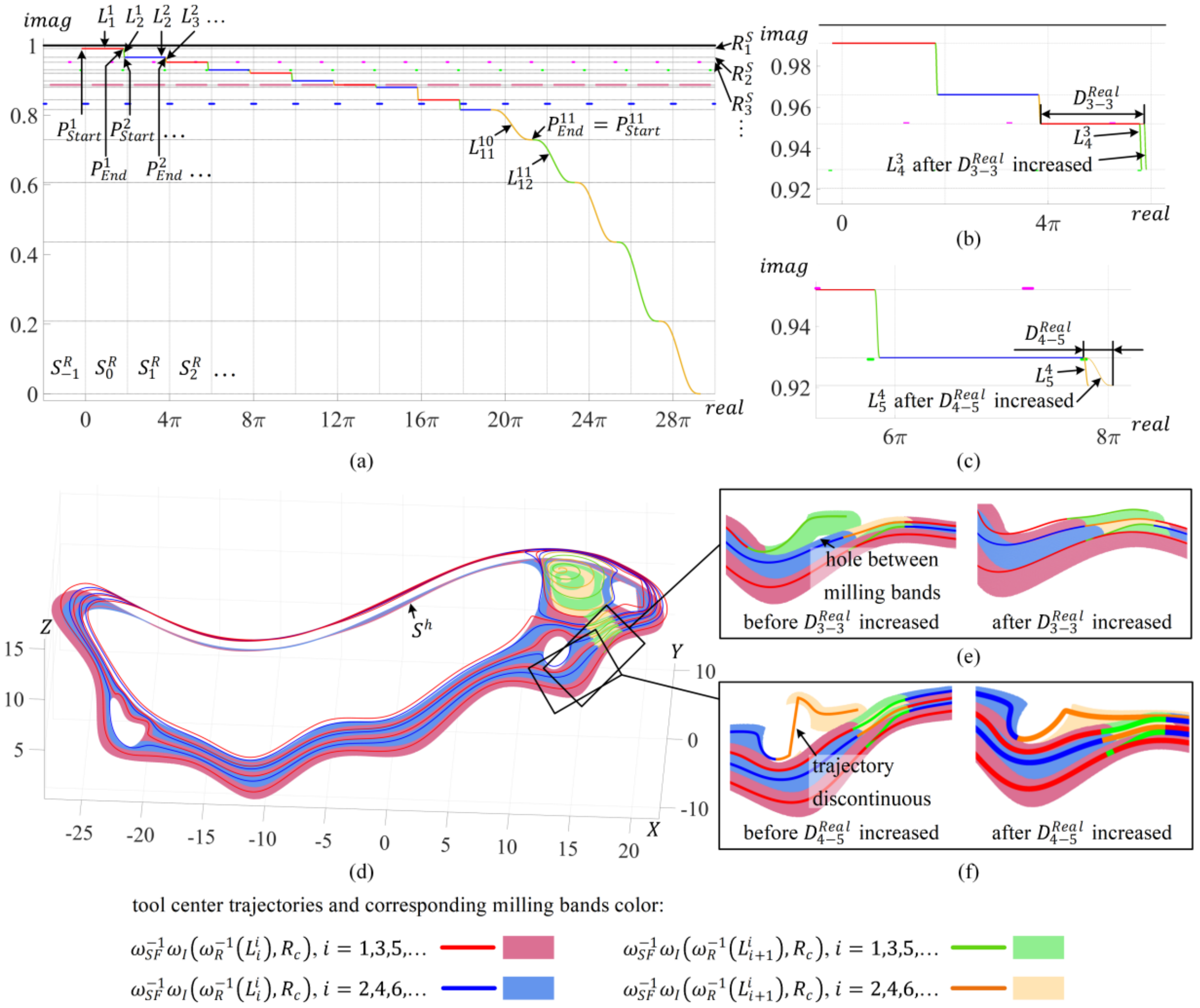

**Fig. 3.** Spiral bridging of iso-parametric trajectories. (a) Bridging trajectory on $S^R$ that avoids linear slits. (d) Spiral tool trajectory obtained by mapping the bridging trajectory on $S^R$ mapping $\omega_R^{-1}\omega_{SF}^{-1}\omega_I(\cdot, K_c)$, along with the corresponding milling band on $S^h$. (b) and (e) Increasing $D_{3-3}^{Real}$ on $S^R$ to eliminate unswept regions between the milling bands of the corresponding spiral trajectories. (c) and (f) Increasing $D_{4-5}^{Real}$ on $S^R$ to prevent the corresponding spiral trajectories from passing through holes on the surface.

Region $S^R$ can be mapped back to $S^S$ via the following mapping:

$$S^S = imag(S^R)e^{i*real(S^R)} \tag{8}$$

Define the mapping $\omega_R^{-1}: S^R \to S^S$, and establish connections among all parallel lines characterized by imaginary components $R_k = \{R_1^S, \ldots, R_k^S, R_{k+1}^S\}$, where $R_{k+1}^S = R_{min}$, within $S^R$. These connections should be formed using smooth curves that bypass straight slits. The corresponding parallel lines are depicted as dashed lines in Fig. 3(a).

Step 1: The endpoints of the trajectory linking two adjacent parallel lines with imaginary components $R_1^S$ and $R_2^S$ are designated as $P_{End}^1$ and $P_{Start}^2$, respectively. These satisfy $imag(P_{End}^1) = R_1^S$, $imag(P_{Start}^2) = R_2^S$, while $real(P_{End}^1)$ remains unspecified within the interval $[0, 2\pi]$. The real component of $P_{Start}^2$ is expressed as $real(P_{Start}^2) = D_{1-2}^{Real} + real(P_{End}^1)$. Define $P_{Start}^1 = P_{End}^1 - 2\pi$, and denote the line segment connecting $P_{Start}^1$ and $P_{End}^1$ as $L_1^1$. Assign $D_{1-2}^{Real} = \frac{\pi}{10}$, $i = 1$. Construct a smooth curve $L_2^1$ connecting$P_{End}^1$ and$P_{Start}^2$ such that it maintains tangency to both parallel lines by refining the function $\sigma(t)$ in Equation A-11 as outlined by She et al. [14].

$$L_{i+1}^i = real\left(P_{End}^i + t\left(P_{Start}^{i+1} - P_{End}^i\right)\right) + i * imag\left(P_{End}^i + \frac{P_{Start}^{i+1} - P_{End}^i}{2\pi}\,\sigma(2\pi t)\right) \quad (9)$$

For $t \in [0, 1]$, when $L_2^1$ intersects a slit on $S^R$, the bridge trajectory $\omega_{SF}^{-1}\omega_I(\omega_R^{-1}(L_2^1), R_c)$ crosses through the gap, resulting in discontinuity. To maintain continuity between bridging trajectories, $D_{1-2}^{Real}$ can be incrementally increased, shifting $L_2^1$ to the right until it bypasses the slit. Once the smallest possible $D_{1-2}^{Real}$ is established, both $P_{End}^1$ and $P_{Start}^2$ are simultaneously determined.

Step 2: The trajectory connecting the parallel lines with imaginary components $R_2^S$ and $R_3^S$ is defined by its endpoints $P_{End}^2$ and $P_{Start}^3$, where $imag(P_{End}^2) = R_2^S$ and $imag(P_{Start}^3) = R_3^S$. The real parts follow $real(P_{End}^2) = D_{2-2}^{Real} + real(P_{Start}^2)$ and $real(P_{Start}^3) = D_{2-3}^{Real} + real(P_{End}^2)$.

To adjust $D_{2-2}^{Real}$, define the segment between $P_{Start}^2$ and $P_{End}^2$ as $L_2^2$, set $D_{2-3}^{Real} = \frac{\pi}{10}$, and $i = 2$. Using Equation 9, connect $P_{Start}^2$ and $P_{End}^3$ to obtain $L_3^2$. Incrementally increase $D_{2-2}^{Real}$ from zero until the milling bands of the ball center trajectory $\omega_{SF}^{-1}\omega_I(\omega_R^{-1}(L_1^0 \cup L_2^1 \cup L_2^2 \cup L_3^2), K_c)$ fully encompass the milling band $BP_2$ on $S^h$. After determining the minimum $D_{2-2}^{Real}$, verify whether $L_3^2$ intersects a slit on $S^R$. If an intersection occurs, gradually increase $D_{2-3}^{Real}$ until $L_3^2$ is clear of the slit. Once the smallest values of $D_{2-2}^{Real}$ and $D_{2-3}^{Real}$ are established, both $P_{End}^2$ and $P_{Start}^3$ are simultaneously defined. Figs. 3(b) and 3(e) depict the gradual increase of $D_{i-i}^{Real}$ for seamless milling band connection, while Figs. 3(c) and 3(f) illustrate the adjustment of $D_{i-i+1}^{Real}$ to prevent trajectory intersections with holes.

By iterating Step 2, the bridging trajectories connecting parallel lines with imaginary components $\{R_1^S, \ldots, R_k^S, R_{k+1}^S\}$ can be effectively adjusted to circumvent slits, where $R_{k+1}^S = R_{min}$. However, as the bridge trajectories approach the center of the spiral, the abrupt turns in the corresponding $C_i^T$ make it challenging to maintain smooth transitions. To mitigate this issue, when the turning radius of $C_i^T$ is small, the initial value of $D_{i-i+1}^{Real}$ is increased from $\frac{\pi}{10}$ to $2\pi$. This modification extends the bridging trajectory, allowing it to form a smoother spiral near the center, as illustrated in Fig. 1(a). The optimal value of $real(P_{Start}^1)$, which minimizes or smooths the bridging trajectory, can be identified through trial and error within the interval $[0, 2\pi]$. The pseudocode for spiral bridging of iso-parametric trajectories is provided in Appendix A.

*2.3. Optimal position of the origin-mapped point in conformal slit mapping*

As outlined in Sections 2.1 and 2.2, generating spiral trajectories via conformal slit mapping necessitates selecting a reference point $O^F$, which maps to the origin $O^S$ and satisfies $O^F \in S^F(\Gamma_0^F) - \Gamma^F$. The placement of $O^F$ plays a pivotal role in ensuring uniform trajectory spacing. For instance, as depicted in Figs. 2(b) and 2(d), positioning $O^F$ at $O_1^F$ (the centroid of $S^F$) results in non-uniform spacing, producing sparser trajectories on the left and denser ones on the right. This imbalance amplifies variations in trajectory scallop height and extends the overall trajectory length (1214.21 in Fig. 2(b) vs. 1013.11 in Fig. 2(d)). Consequently, determining the optimal placement of $O^F$ is essential for minimizing scallop height variations. This necessitates formulating an optimization criterion for $O^F$ and identifying the position that minimizes this criterion.

2.3.1. Establish evaluation criteria for the origin-mapped point position

Utilizing the properties of conformal slit mapping, a scalar field $T$ is constructed on $S$ such that $|dT| \neq 0$ on $S - \Gamma$, where $dT$ represents the gradient of the scalar field $T$. For any point $P^S \in S^S$, with a distance $x(P^S) \in [R_{min}, 1]$ from $O^S$, the scalar value at $P^S$ is given by:

$$T^S(P^S) = f(x) \tag{10}$$

where $f$ is an unknown monotonically increasing function satisfying:

$$f(R_{min}) = 0 \text{ and } f'(x) > 0 \tag{11}$$

Since the meshes of $S^S$ and $S$ are topologically equivalent, the scalar field $T^S$ on $S^S$ can be mapped equivalently onto $S$ to obtain $T$. Thus, the function $T$ on $S$ is ultimately determined by the unknown function $f$.

Let $CL_i = \{P \in S | T(P) = T_i\}$ and $CL_{i+1} = \{P \in S | T(P) = T_{i+1}\}$ represent the adjacent isocurves of $T$ on $S$. If two machining points $P_i$ and $P_{i+1}$, located on these adjacent isocurves, satisfy:

$$real\left(exp\left(\omega_{FS}(P_i, O^F)\right)\right) = real\left(exp\left(\omega_{FS}(P_{i+1}, O^F)\right)\right) \tag{12}$$

i.e., $\omega_{FS}(P_i, O^F)$ and $\omega_{FS}(P_{i+1}, O^F)$ belong to the same meridian on $S^S$, then $P_{i+1}$ is considered an adjacent machining point to $P_i$.

The machining scallop height between $P_i$ and $P_{i+1}$ can be determined using the following equation [31]:

$$h = \frac{K_s + K_c}{8} \left\|\overrightarrow{P_i P_{i+1}}\right\|_2^2 + o\left(\left\|\overrightarrow{P_i P_{i+1}}\right\|_2^3\right) \tag{13}$$

$K_c$ represents the ball-end mill radius, $K_s$ denotes the normal curvature of the surface $S$ at $P_i$ in the direction of $\overrightarrow{P_i P_{i+1}}$, and $\overrightarrow{P_i P_{i+1}}$ is the vector between $P_i$ and $P_{i+1}$. $o\left(\left\|\overrightarrow{P_i P_{i+1}}\right\|_2^3\right)$ is the cubic infinitesimal of its Euclidean norm.

When $P_{i+1}$ is near $P_i$, the vector $\overrightarrow{P_i P_{i+1}}$ aligns with the gradient $\nabla T(P_i)$. Taylor's theorem is then applied to derive the following equation:

$$(\|\nabla T(P_i)\|_2)\left(\left\|\overrightarrow{P_iP_{i+1}}\right\|_2\right) + o\left(\left\|\overrightarrow{P_iP_{i+1}}\right\|_2^2\right) = |T_{i+1} - T_i| \tag{14}$$

Neglecting higher-order infinitesimals and combining Equations (13) and (14) yields:

$$h = |T_{i+1} - T_i|^2 \frac{K_s + K_c}{8\|\nabla T(P_i)\|_2^2} \tag{15}$$

The relationship among $CL_i$, $CL_{i+1}$, $h$, $\nabla T(P_i)$, $K_s$ and $K_c$ is illustrated in Fig. 4.

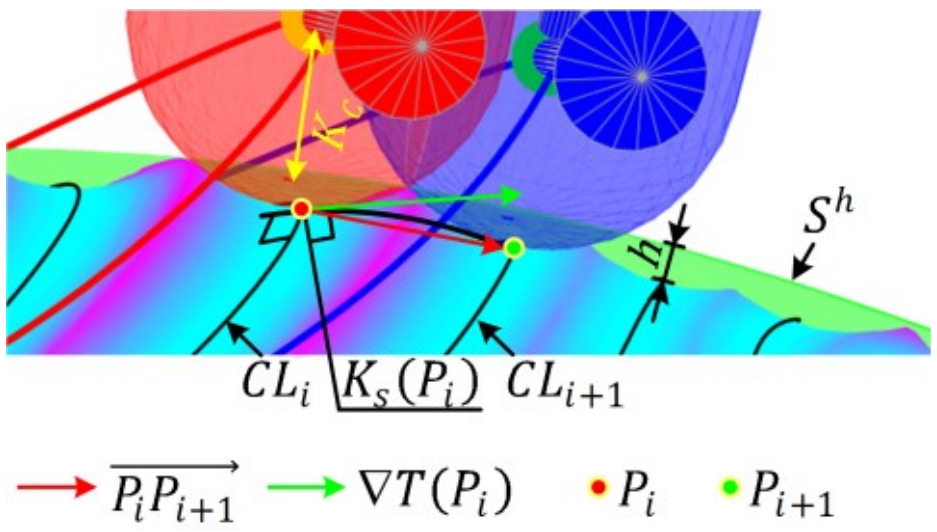

**Fig. 4.** Relationship between $CL_i$, $CL_{i+1}$, $h$, $\nabla T(P_i)$, $K_s$ and $K_c$.

Equation (15) provides a surface-local scallop height evaluation measure that excludes the spacing between adjacent tool paths, ensuring it remains unaffected by abrupt increases in actual tool path spacing, as observed at points $P_4$ and $P_5$ in Fig. 2. Since $|T_{i+1} - T_i|^2$ remains constant, Equation (15) implies that the variation in machining height between adjacent isocurves can be characterized by the changes in $\frac{K_s+K_c}{8\|\nabla T\|_2^2}$. The average value $Avg$ of $\frac{K_s+K_c}{8\|\nabla T\|_2^2}$ over $S - \Gamma$ is given by:

$$Avg = \frac{1}{A_S}\int_{S-\Gamma}\left(\frac{K_s+K_c}{8\|\nabla T\|_2^2}\right) dS \tag{16}$$

where $A_S$ represents the area of $S$, and $dS$ denotes the area element. The global fluctuations of $\frac{K_s+K_c}{8\|\nabla T\|_2^2}$ around $Avg$ can be evaluated using the following symmetric energy function:

$$E^S = \int_{S-\Gamma}\left(\frac{K_s+K_c}{8\|\nabla T\|_2^2} + \frac{8\|\nabla T\|_2^2}{K_s+K_c}\right) dS \tag{17}$$

In Appendix B, it is demonstrated that when $\frac{K_s+K_c}{8\|\nabla T\|_2^2} = Avg$ on $S - \Gamma$, the expression $E^S$ in Equation (17) reaches its minimum.

For a given surface, the quantity $E^S$ is determined by $T$, which is influenced by the unknown function $f$ and the position of $O^F$ on $S^F(\Gamma_0^F) - \Gamma^F$. To assess the quality of a given position of $O^F$, the function $f$ that minimizes $E^S$ must be determined. This requires solving the following functional problem:

$$E^S_{min}(O^F) = min_{f\in\{f|f(R_{min})=0, f'(D^S)>0 \text{ for } D^S\in[R_{min},1]\}} E_S(f, O^F) \tag{18}$$

Appendix C introduces a method for determining the function $f$ that minimizes $E^S$ and includes the pseudocode for the algorithm. The computational complexity of this algorithm is $O\left(N_F log\left(\frac{1}{E^S_\varepsilon}\right)\right)$, where $N_F$ represents the number of triangular faces of $S$, and $E^S_\varepsilon$ denotes the convergence tolerance error of the energy $E^S$.

2.3.2. Optimization of the origin-mapped point position

With the ability to efficiently compute $E^S_{min}(O^F)$ for $O^F \in S^F(\Gamma^F_0) - \Gamma^F$, the next step involves determining the optimal position of $O^F$ that minimizes $E^S_{min}(O^F)$:

$$O^F_{opt} = argmin_{O^F\in S^F(\Gamma^F_0)-\Gamma^F} E^S_{min}(O^F) \tag{19}$$

As illustrated in Fig. (5), $E^S_{min}(O^F)$ was computed for 3,225 discrete points sampled from $S^F(\Gamma^F_0) - \Gamma^F$. Each computation required an average of 1.53 seconds, leading to a total processing time of approximately 1.37 hours. All calculations in this study were performed on a desktop computer equipped with an Intel i5-10400F CPU and an NVIDIA GeForce GTX 1600 GPU. Consequently, employing a traversal-based approach to determine the optimal $O^F$ may present considerable challenges for users with constrained computational resources.

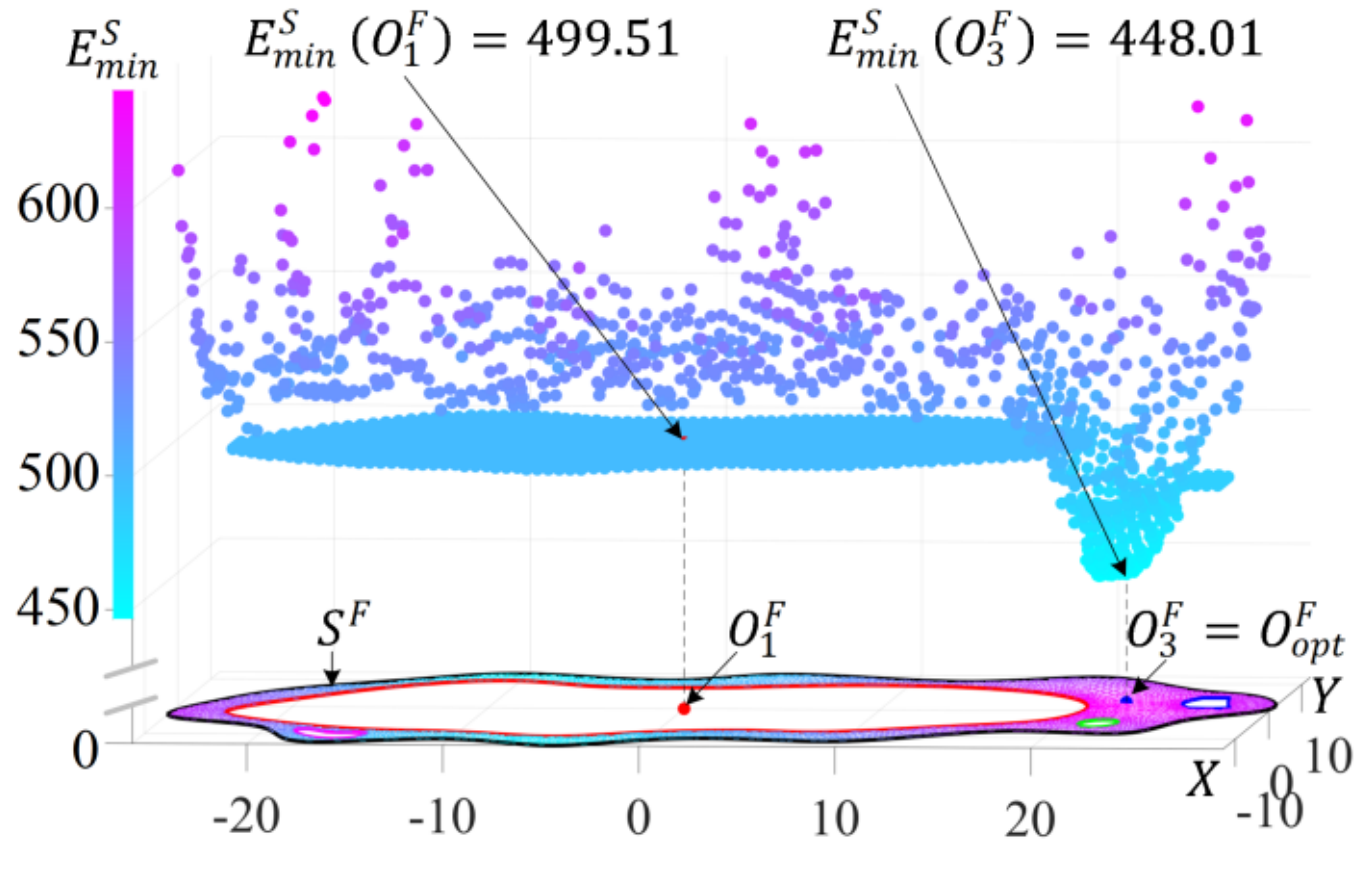

**Fig. 5.** Results of the traversal calculation of $E^S_{min}$ on Surface $S^F$.

Based on the distribution of $E^S_{min}$ over the region $S^F(\Gamma^F_0) - \Gamma^F$, it is observed that $E^S_{min}$ exhibits smoothness and convexity in the vicinity of the optimal point $O^F_{opt}$. Furthermore, within the regions $S^F(\Gamma^F_i)$ for $i = 1,2,\ldots,m$, $E^S_{min}$ remains gradient-free. This occurs because positional variations of $O^F$ within a specific $S^F(\Gamma^F_i)$ do not influence the outcomes of the conformal slit mapping calculation. As depicted in Figs. 1(b) and 1(d), positioning $O^F$ at $O^F_1$ or $O^F_5$ produces identical computational results. Consequently, the optimal $O^F_{opt}$ is determined using the following modified gradient descent method:

Select any point on $S^F$ as the initial point $O^F$ for iteration. The gradient of $E^S_{min}$ at $O^F$ can be determined using the following numerical method:

$$\nabla E^S_{min}(O^F) = \left(E^S_{min}(O^F + \varepsilon\vec{u}) - E^S_{min}(O^F)\right)\frac{\vec{u}}{\varepsilon} + \left(E^S_{min}(O^F + \varepsilon(\vec{u})^\perp) - E^S_{min}(O^F)\right)\frac{(\vec{u})^\perp}{\varepsilon} \tag{20}$$

In this expression, $\vec{u}$ and $(\vec{u})^\perp$ are any two orthogonal unit vectors on $S^F$. The expressions $O^F + \varepsilon\vec{u}$ and $O^F + \varepsilon(\vec{u})^\perp$ denote the positions obtained by moving $O^F$ a small distance $\varepsilon$ along $\vec{u}$ and $(\vec{u})^\perp$, respectively. Define the iterative update as:

$$O^F = O^F + \frac{S^F_{min}}{100\left\|\nabla E^S_{min}(O^F)\right\|_2 (1.01)^{Idx}} \nabla E^S_{min}(O^F) \tag{21}$$

where $S^F_{min}$ represents the radius of the largest inscribed circle in $S^F(\Gamma^F_0)$. The iteration continues until $E^S_{min}$ increases with respect to the position of $O^F$.

To prevent the gradient $\nabla E^S_{min}(O^F)$ from vanishing within the region $S^F(\Gamma^F_i)$ for $i = 1,2,\ldots,m$, which would halt the iteration process, if $O^F$ falls within $S^F(\Gamma^F_i)$ for $i = 1,2,\ldots,m$, the iteration position of $O^F$ is shifted to the point $O^F_{i\text{-}off}$, and the iteration process is restarted. The point $O^F_{i\text{-}off}$ is defined as:

$$O^F_{i\text{-}off} = argmin_{O^F \in \Gamma^F_{i\text{-}off}} E^S_{min}(O^F) \tag{22}$$

where $\Gamma^F_{i\text{-}off}$ is the curve obtained by offsetting $\Gamma^F_i$ outward and is a subset of $S^F$. As illustrated in Fig. 6(a), the initial starting point $O^F_{start-1}$ is selected to begin the iterative process for determining $O^F_{opt}$. When $O^F_{start-1}$, following the gradient of $E^S_{min}$, enters $S^F(\Gamma^F_1)$, it is transferred to $O^F_{1-off}$ on $\Gamma^F_{1-off}$, where $E^S_{min}$ attains a minimum, and the iteration resumes.

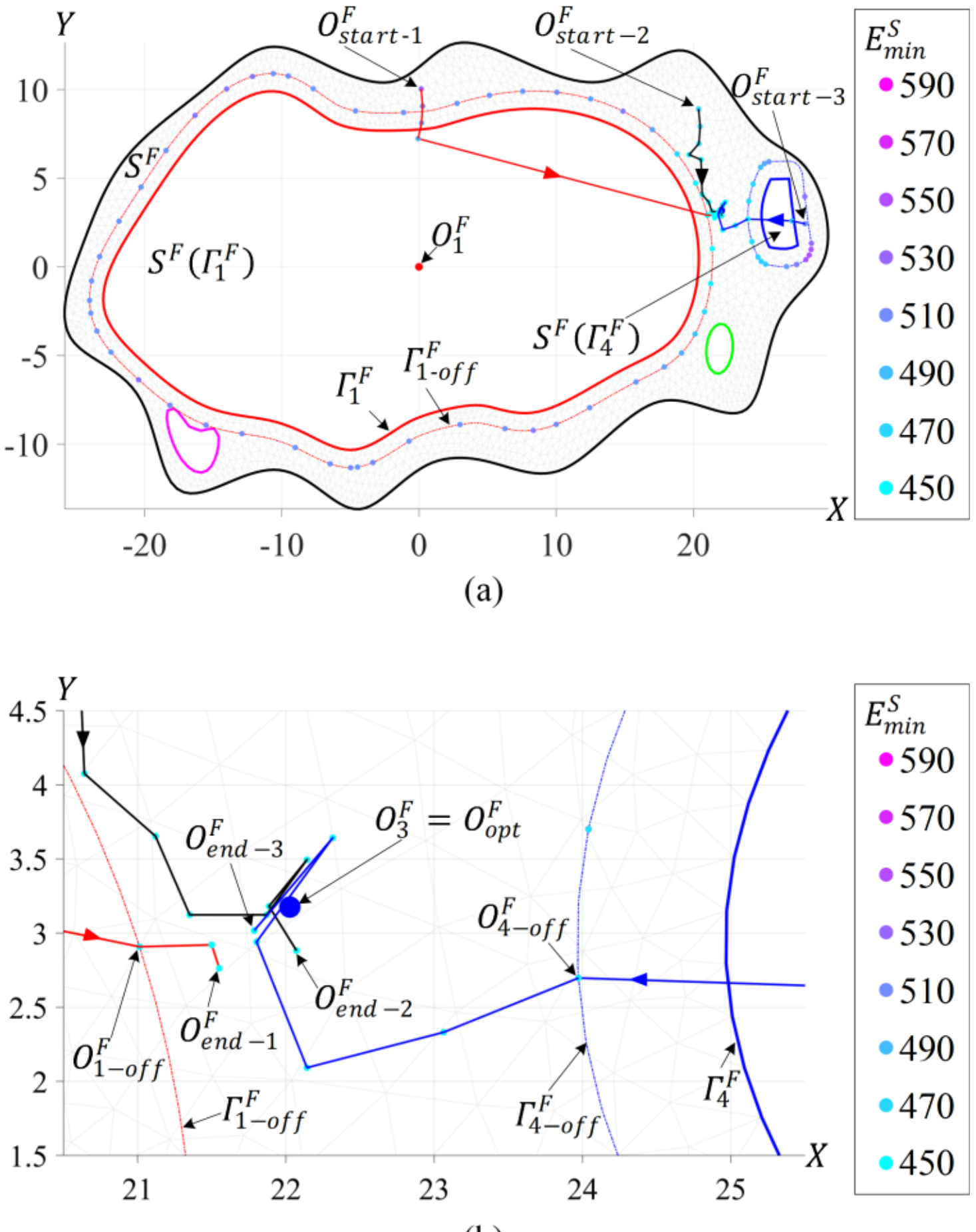

**Fig. 6.** Optimization of the origin-mapped point for a complex asymmetric surface.

Fig. 6(b) depicts the iterative paths originating from $O^F_{start-1}$, $O^F_{start-2}$, and $O^F_{start-3}$, along with their respective iteration endpoints $O^F_{end-1}$, $O^F_{end-2}$, and $O^F_{end-3}$. As observed, $O^F_{end-1}$, $O^F_{end-2}$, and $O^F_{end-3}$ are all positioned near the optimal point $O^F_{opt}$ identified through traversal. However, the three iterative paths necessitate only 67, 33, and 38 evaluations of $E^S_{min}$, with total computation times of 104.98s, 56.29s, and 60.32s, respectively. These durations are markedly shorter than the 3225 evaluations and 1.37 hours required by the traversal method.

In Appendix C, the pseudocode for locating the optimal position of $O^F$ is provided.

## 3. Experiment and analysis

### *3.1. Numerical arithmetic experiments*

Figs. 7(b) and 7(c) compare conventional machining approaches with the proposed method for processing the surface depicted in Fig. 7(a). Conventional techniques segment surfaces containing holes into inner and outer regions using annular and disk mappings, while the proposed approach eliminates this need, generating tool paths without artificial boundaries. This results in shorter (294.24 vs. 245.47), smoother, and more uniform spaced trajectories.

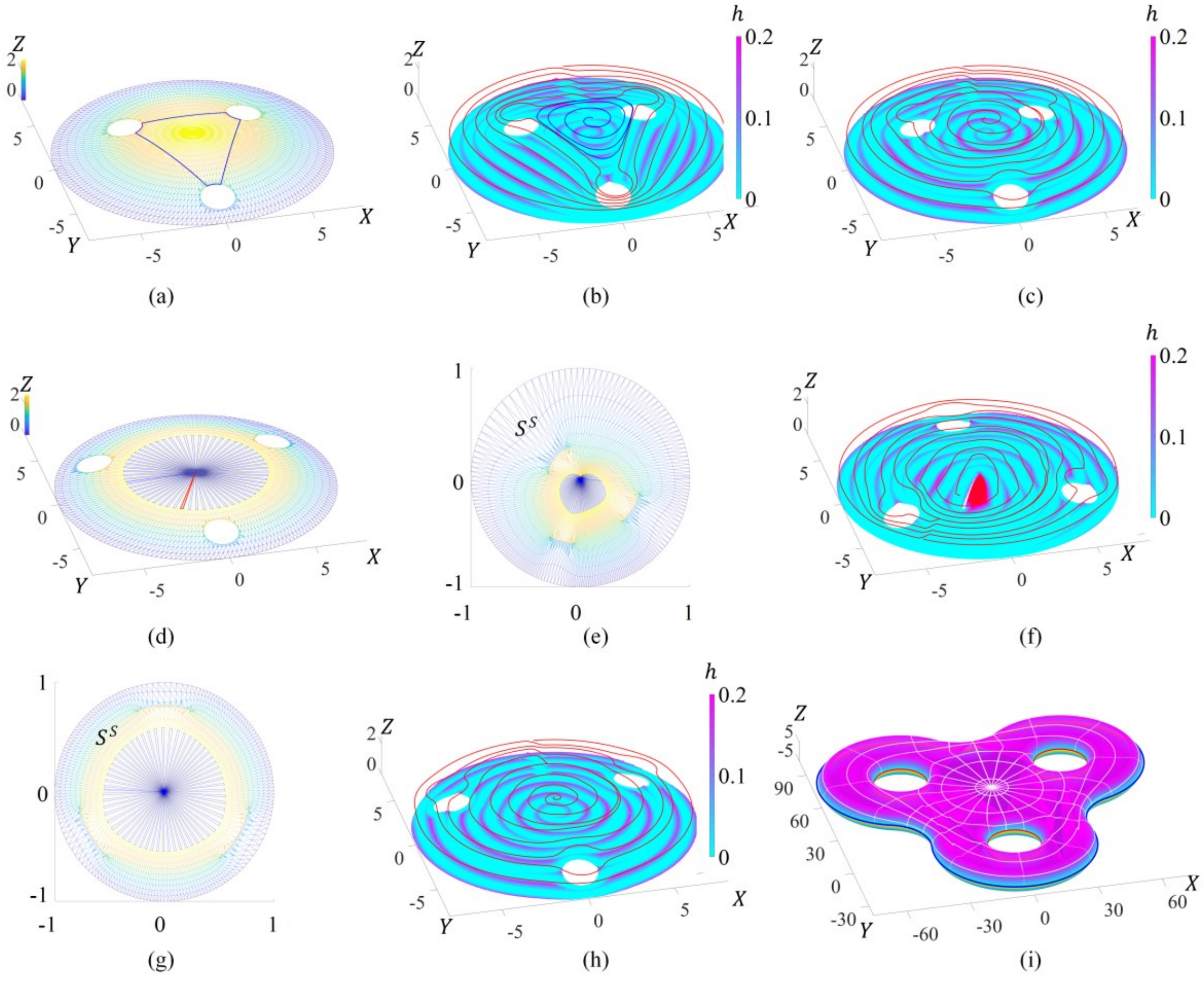

**Fig. 7.** Comparison of milling trajectory generation using various methods, conformal flattening of low-quality meshes, and parameterization of high-genus surfaces.

Fig. 7(d) presents a low-quality mesh with elongated triangular facets at the center. A comparison between the CSM method [13] and the proposed approach reveals that the CSM method maps these facets to the center, causing severe iso-parametric deformation. The proposed method avoids removing the triangular mesh, ensuring a higher conformality. Consequently, it produces shorter (284.26 vs. 226.11), smoother, and more uniform spaced trajectories.

Fig. 7(i) demonstrates the algorithm's capability in parameterizing high-genus surfaces. Cutting along the red tunnel loop transforms a genus-3 surface into a surface with six holes, enabling the definition of a South Pole and a North Pole. Employing conformal slit mapping, the South and North Poles are projected to the center and the outer boundary of the unit disk, respectively. This approach facilitates binary parameterization, while spiraling iso-parametric lines allow monomial parameterization. The method offers promising applications in manufacturing complex models, including 3D printing.

Table 1 presents the computational time of the path spacing control algorithm under various parameter conditions for controlling trajectory spacing, where $t_1$ represents the computation time for each case. The value $STR_1 = \frac{k}{t_1} N_S log(N_C) log\left(\frac{1}{\varepsilon}\right)$ represents the ratio of algorithm complexity to actual computation time, with $k$ being the number of generated trajectories.

**Table 1.** The computational time of the path spacing control algorithm under different parameter conditions.

| Case Number | $h$ | $k$ | $N_S$ | $N_C$ | $\varepsilon$ | $t_1$ (s) | $STR_1$ |
|---|---|---|---|---|---|---|---|
| 1.1 | 0.5 | 17 | 79842 | 1000 | 0.01 | 57.19 | $7.55\times10^5$ |
| 1.2 | 0.2 | 25 | 79842 | 1000 | 0.01 | 43.80 | $1.45\times10^6$ |
| 1.3 | 0.1 | 34 | 79842 | 1000 | 0.01 | 61.93 | $1.39\times10^6$ |
| 1.4 | 0.1 | — | 79842 | 100 | 0.01 | — | — |
| 1.5 | 0.1 | 28 | 8377 | 1000 | 0.01 | 16.46 | $4.53\times10^5$ |
| 1.6 | 0.1 | 34 | 79842 | 10000 | 0.01 | 319.03 | $3.61\times10^5$ |
| 1.7 | 0.1 | 34 | 320251 | 1000 | 0.01 | 613.51 | $5.65\times10^5$ |
| 1.8 | 0.1 | 34 | 79842 | 1000 | 0.1 | 11.42 | $3.78\times10^6$ |

Table 1 indicates that the value of $STR_1 \in [3.61 \times 10^5, 3.78 \times 10^6]$ remains within a single order of magnitude, demonstrating the algorithm's reliability in controlling path spacing. Figs. 8(a)- 8(c) correspond to Cases 1.1-1.3, demonstrating that the algorithm maintains the present maximum milling scallop height on complex surfaces. In Case 1.4, sparse trajectory points cause computational errors in milling bands. Case 1.5 shows overly sparse discrete points, resulting in a small $k$-value and excessively large trajectory spacing. Fig. 8(f) corresponds to Case 1.6, where increasing $N_S$ and $N_C$ to sufficiently high values yield the correct computational result, but at the cost of a significantly higher computational burden.

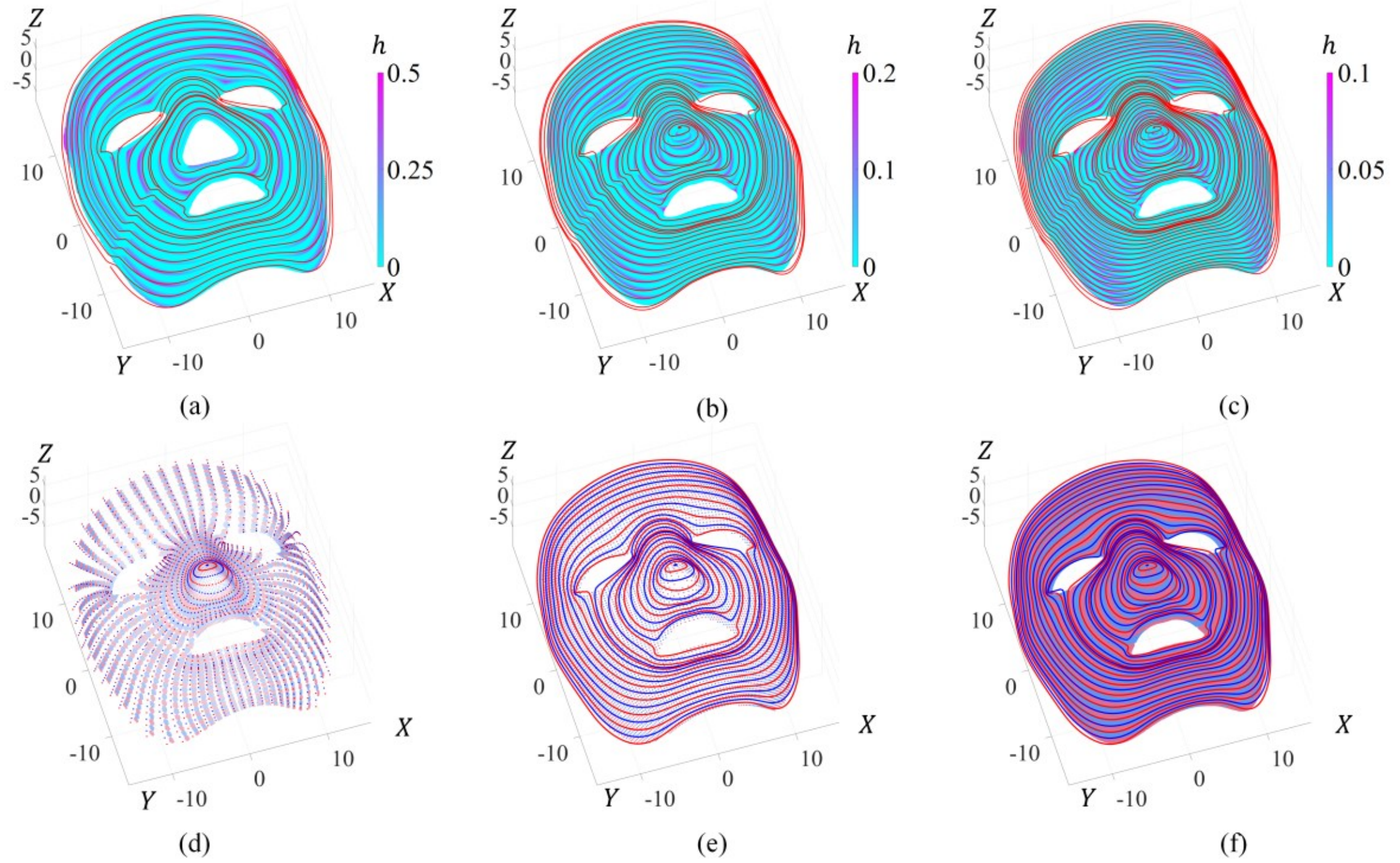

**Fig. 8.** Milling trajectories and states obtained using the parameters set in Table 1.

Table 2 compares the computational time of the algorithm under different parameter conditions with the computed $E^S_{min}$.

**Table 2.** Computational time for calculating $E^S_{min}$ under different parameter conditions.

| **Case Number** | $N_F$ | $E^S_\varepsilon$ | $t_2$ **(s)** | $E^S_{min}$ | $STR_2$ |
|---|---|---|---|---|---|
| 2.1 | 1406 | 0.1 | 0.31 | 452.14 | $1.04\times10^4$ |
| 2.2 | 2033 | 0.1 | 1.15 | 451.14 | $4.07\times10^3$ |
| 2.3 | 5039 | 0.1 | 1.85 | 450.66 | $6.27\times10^3$ |
| 2.4 | 8646 | 0.1 | 4.1 | 453.61 | $4.85\times10^3$ |
| 2.5 | 8646 | 0.01 | 7.54 | 453.99 | $5.28\times10^3$ |
| 2.6 | 8646 | 0.001 | 8.61 | 453.99 | $6.93\times10^3$ |

The term $t_2$ represents the computation time for each case. The value $STR_2 = \frac{N_F}{t_2} log\left(\frac{1}{E^S_\varepsilon}\right)$ represents the ratio of algorithm complexity to actual computational time.

Table 2 shows that $STR_2 \in [4.07 \times 10^3, 1.04 \times 10^4]$ remains stable within a single order of magnitude, demonstrating the reliability of the algorithm complexity analysis for calculating $E_{min}^S$. Case 2.2 achieves significantly faster computation, requiring only 1.86% of the time used in Case 1.3 (1.15s vs. 61.93s). This efficiency may originate from specific reasons:

Even with a coarser grid, the computed results remain accurate. Table 2 shows that varying surface discretization from 1406 to 8646 results in less than a 1% difference in $E_{min}^S$ $\left(max(E_{min}^S) = 453.99 \text{ vs } min(E_{min}^S) = 450.66\right)$. However, the trajectory spacing control algorithm requires dense point sampling on the iso-scallop height surface and tool trajectories. Insufficient sampling leads to inaccuracies, while excessive discrete points significantly increase computational costs.

The computational complexity of calculating $E_{min}^S$ is $O\left(N_F log\left(\frac{1}{E_\varepsilon^S}\right)\right)$, primarily influenced by $N_F$. In contrast, the trajectory spacing control algorithm has a complexity of $O\left(kN_S log(N_C) log\left(\frac{1}{\varepsilon}\right)\right)$, involving two linear growth terms: $k$ and $N_S$. When scallop height requirements decrease, is smaller, increasing trajectory density and $k$, the efficiency of calculating $E_{min}^S$ surpasses that of the trajectory spacing control algorithm.

Fig. 9(a) illustrates a centrally symmetric surface with symmetry centered at point $O$. Using symmetry, the optimal mapped origin position is $O_{opt}^F = \omega_F(O)$. As shown in Fig. 9(c), iterative searches from different initial points consistently converge to the vicinity of $O_{opt}^F$, demonstrating the effectiveness of the optimization process.

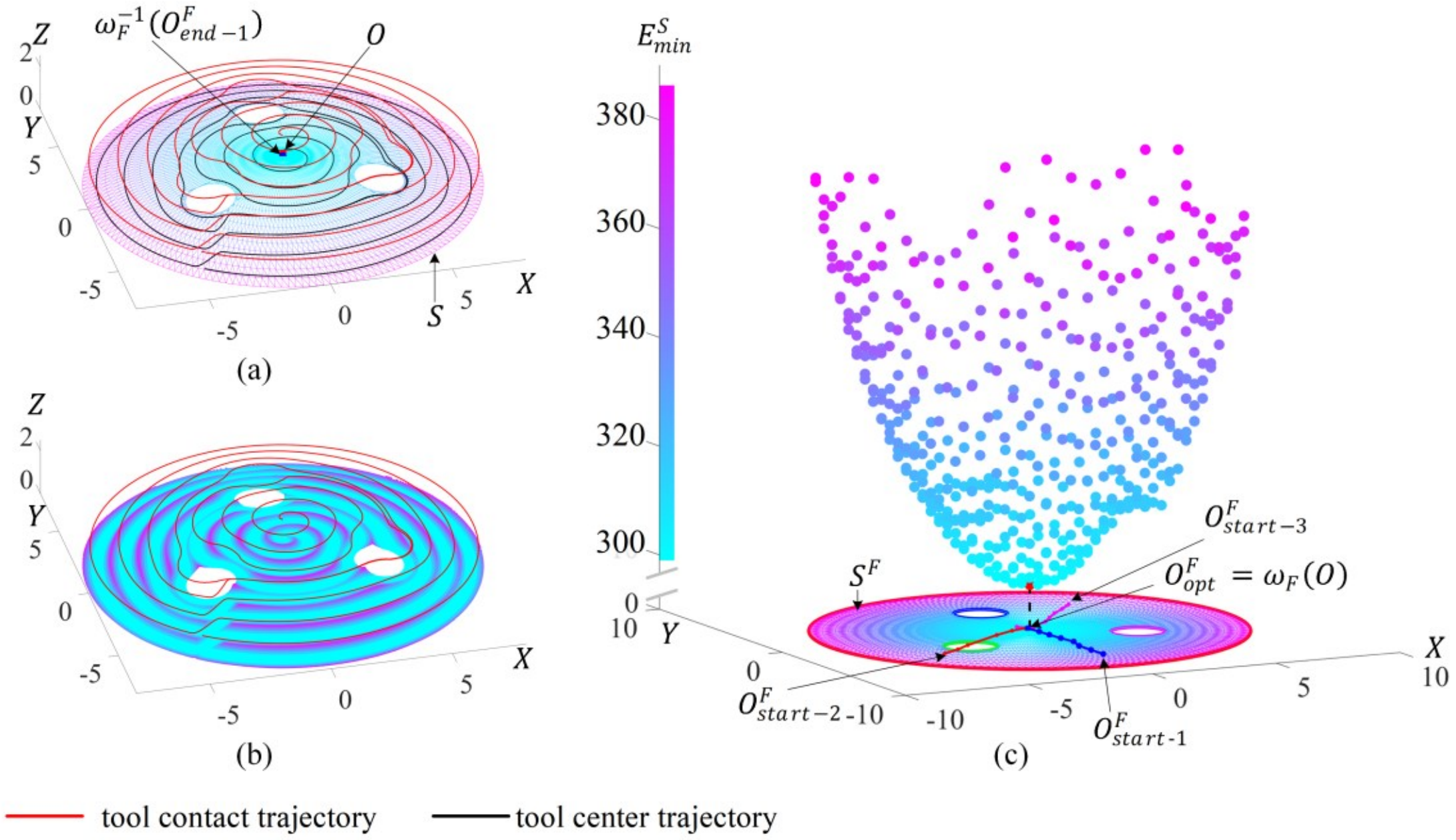

**Fig. 9.** Optimization of the origin-mapped point for a centrally symmetric surface and its effect on trajectory generation. (a) Centrally symmetric surface with symmetry center $O$. (b) Generated trajectory based on the optimized origin position $\omega_F^{-1}(O_{end-1}^F)$. (c) Convergence of iterative searches to $O_{opt}^F$ from different initial points on the surface.

### 3.2. Experiment on surface milling

Fig. 10(a) illustrates the IRB6600 robotic arm milling platform, with a PCB triaxial accelerometer mounted on the platform spindle, operating at a sampling rate of 2000 Hz. Fig. 10(b) shows the three-coordinate laser measurement platform, which boasts a measurement accuracy of 20 μm.

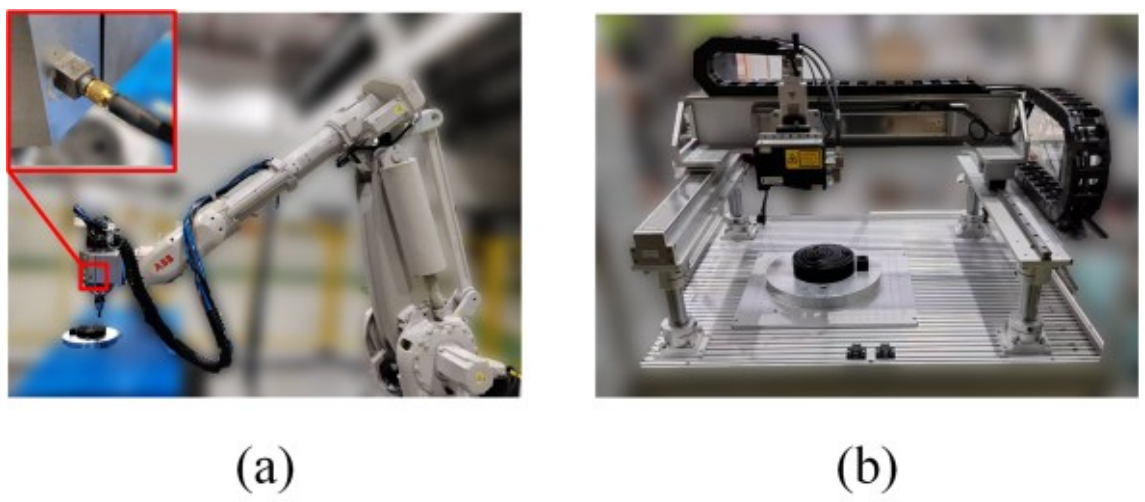

**Fig. 10.** Experimental platforms: (a) Machining experiment platform, (b) Measurement experiment platform.

Machining was performed using traditional trajectory 1 from Fig. 7(b) and proposed trajectory 2 from Fig. 7(c), producing workpieces 1 and 2. Their configurations and surface morphologies are presented in Fig. 11. The machining error range was set to $[0\text{cm}, 0.2\text{cm}]$. Workpiece 1 had an actual error range of $[-0.021\text{cm}, 0.216\text{cm}]$, corresponding to an error ratio of $[-0.5\%, 8.0\%]$. Workpiece 2 exhibited an actual machining error range of $[0.013\text{cm}, 0.224\text{cm}]$, with an error ratio of $[0.0\%, 12.0\%]$. The machining scallop height error for both workpieces remained below 12%, validating the effectiveness of the proposed trajectory planning algorithm in maintaining consistent scallop height control.

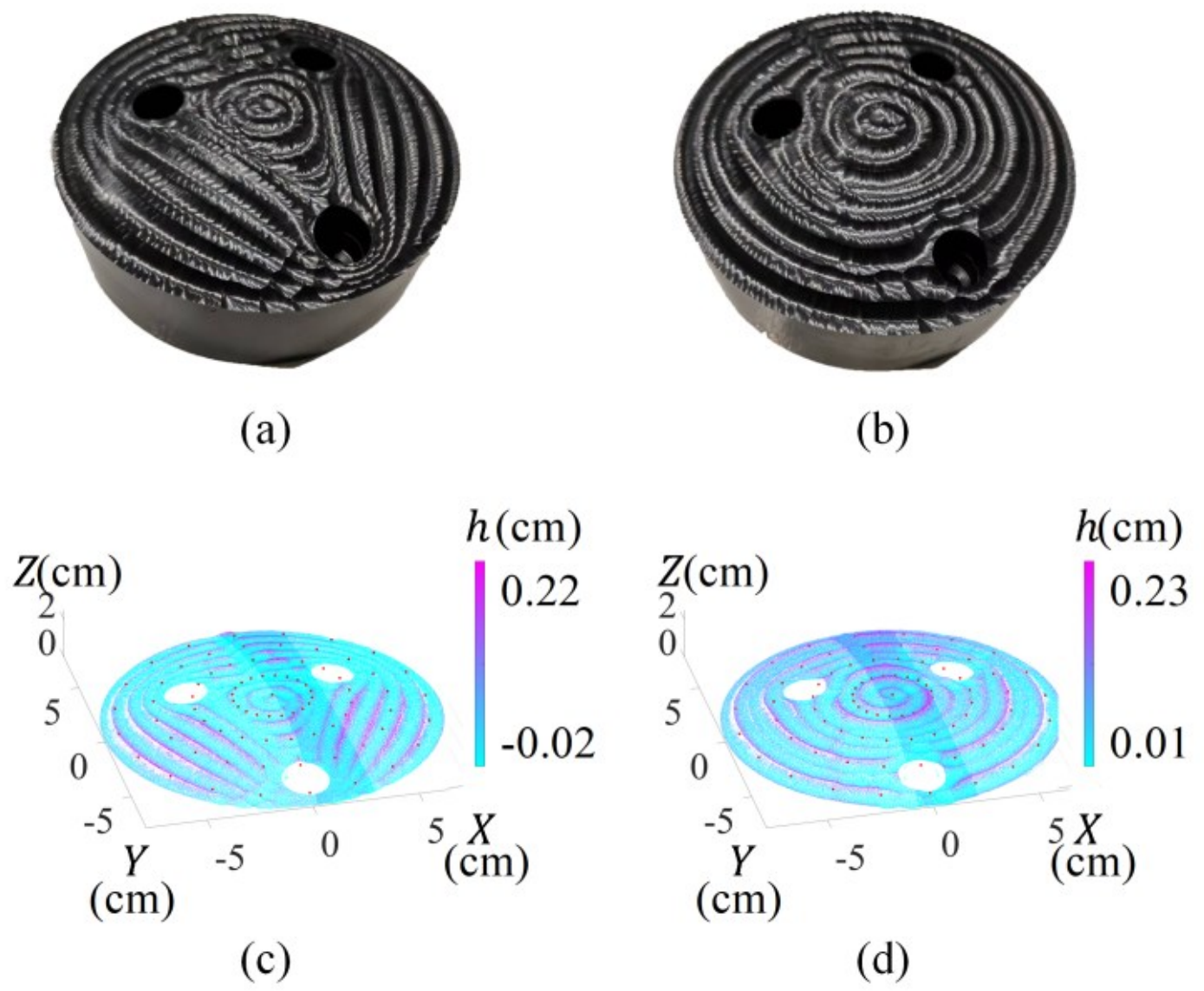

**Fig. 11.** Workpieces and scanned scallop height distribution, with red points indicating maximum scallop height sampling locations. (a) and (c) correspond to Workpiece 1, while (b) and (d) correspond to Workpiece 2.

Let the red sampling points in Figs. 11(c) and 11(d) be denoted as $\{sp_1, sp_2, \ldots, sp_N\}$. The spatial uniformity of scallop height distribution was assessed using the maximum scallop variance index $V_{max}$, defined as follows:

$$V_{max} = \frac{1}{N}\sum_{i=1}^{N}\left(h_{max}(sp_i, R_c) - \overline{h_{max}}\right)^2 \quad (23)$$

In this context, $h_{max}(sp_i, R_c)$ refers to the highest scallop height within a localized region around $sp_i$, defined by a radius $R_c$. Meanwhile, $\overline{h_{max}}$ is the average value of the set $\{h_{max}(sp_1, R_c), h_{max}(sp_2, R_c), \dots, h_{max}(sp_N, R_c)\}$ . When comparing Workpiece 2 to Workpiece 1, the value of $V_{max}$ for Workpiece 2 dropped by 15.63% (from 0.160 to 0.135), which suggests that the new trajectory planning approach significantly enhanced the consistency of scallop height across the surface.

Fig. 12(a) and 12(b) present the total acceleration signals recorded during the machining of Workpiece 1 and Workpiece 2, respectively. Upon analyzing these signals, it was found that machining time, average spindle impact, and spindle impact variance decreased by 7.36% (from 242.20s to 224.38s), 27.79% (from 1.21× $10^{-4} m/s^2$ to 8.74× $10^{-5} m/s^2$), and 55.98% (from 3.43× $10^{-8}$ to 1.51× $10^{-8}$), respectively.

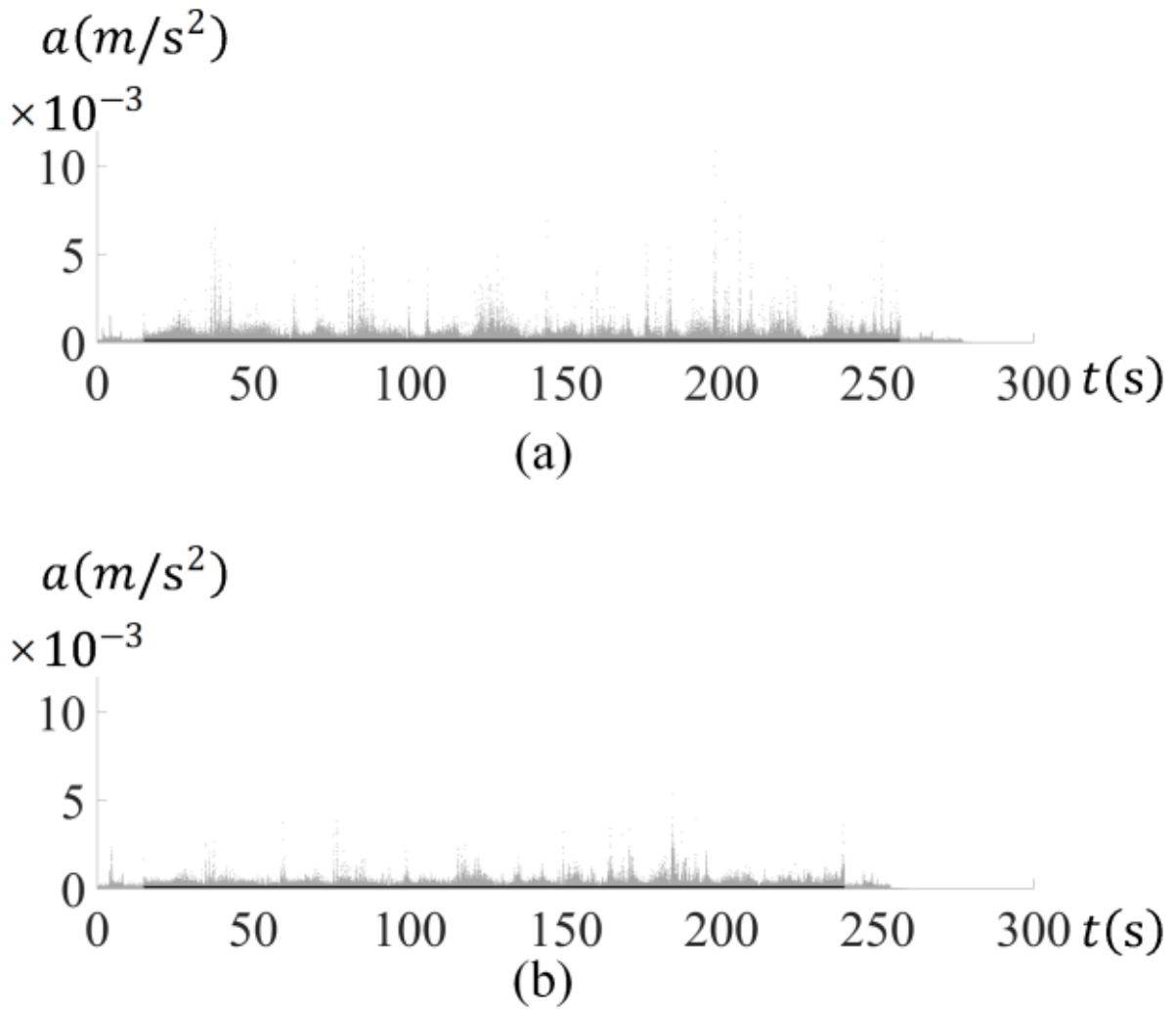

**Fig. 12.** Total acceleration of the Spindle: (a) During machining of Workpiece 1, (b) During machining of Workpiece 2.

## 4. Summary and outlook

### *4.1. Summary*

The conformal slit mapping-based two-dimensional spiral complete coverage trajectory planning algorithm is applied to the ball-end milling of complex surfaces. Unlike conventional methods that rely on decomposing the surface into subregions, this approach removes the need for boundary segmentation, allowing the tool paths to seamlessly navigate around subregion boundaries without unnecessary sharp turns or discontinuities. Consequently, the resulting trajectories are shorter, smoother, and have more consistent spacing.

In this algorithm, the proposed trajectory spacing control method effectively manages the maximum machining scallop height, keeping it within the range of -0.5% to 12%, while minimizing trajectory redundancy. The bridging technique connects the trajectories into a continuous spiral path, avoiding surface holes, thus eliminating unnecessary tool lifts and ensuring smooth motion. During trajectory generation, a reference point $O^F$ is selected on the surface flattening plane and mapped to the origin using conformal slit mapping. To enhance computational efficiency, we present a method to determine the optimal position of $O^F$ without the need for trial-and-error with various candidate points. Instead, a functional energy model is used to analyze trajectory spacing uniformity, and the optimal position for $O^F$ is found by searching along the energy gradient. Simulations demonstrate that a single energy computation takes only 1.86% of the total trajectory generation time, and gradient-based optimization reduces the number of computations for finding the optimal $O^F$ position by 2.08% compared to a traversal search. Consequently, the optimization time for $O^F$ is reduced to just 2.12% of the time required by traversal search.

Milling experiments comparing the proposed algorithm with the traditional method show that, in addition to enhancing the uniformity of the residual height distribution by 15.63%, the developed approach also decreased machining time, average spindle impact, and spindle impact variance by 7.36%, 27.79%, and 55.98%, respectively.

*4.2. Outlook*

In addition to planning tool paths for ball-end milling, the proposed algorithm offers valuable insights for applications such as non-spherical tool milling, surface polishing, and three-dimensional printing. For high-genus surfaces, we present a novel parameterization method in the case study of Fig. 7(i), which helps minimize parameterization singularities.

Effectively cutting and defining the north and south poles of the surface to achieve uniform parameterization is a complex higher-dimensional search problem. The optimization technique used to determine the optimal position of $O^F$ in our approach provides a helpful reference for speeding up this process.

**Acknowledgments**

This study was supported by the National Natural Science Foundation of China (52188102, 52375495). The author Shen. C expresses thanks to Professor Zhang. X for their profound guidance throughout this research. The author also thanks Dr. Xu. B for their support in the experiments and expresses gratitude to David. Gu for offering the summer online course on computational conformal geometry. Finally, the author would like to once again thank MMS. Nasser for their previous assistance.

**Funding**

This research received no specific grant from any funding agency in the public, commercial, or not-for-profit sectors.

**Declaration of conflicting interests**

The Authors declare that there is no conflict of interest.

**References**

[1] R.A. Mali, T.V.K. Gupta, J. Ramkumar, A comprehensive review of free-form surface milling– Advances over a decade. J. Manuf. Process. 62 (2021) 132–167. https://doi.org/10.1016/j.jmapro.2020.12.014

[2] G. Wang, L. Zhao, Q. Liu, X. Li, Y. Sun, M. Chen, Modelling and experimental investigation of micro-dimpled structures milling with spiral trajectory tool reciprocating motion. Chin. J. Aeronautics 38 (2025) 102990. https://doi.org/https://doi.org/10.1016/j.cja.2024.03.027

[3] W. Xiao, G. Liu, G. Zhao, Generating the tool path directly with point cloud for aero-engine blades repair. Proceedings of the Institution of Mechanical Engineers, Part B: Journal of Engineering

Manufacture 235 (2021) 877–886. https://doi.org/10.1177/0954405420970915

[4] E.U. Acar, H. Choset, Y. Zhang, M. Schervish, Path Planning for Robotic Demining: Robust Sensor-Based Coverage of Unstructured Environments and Probabilistic Methods. Int. J. Rob. Res. 22 (2003) 441–466. https://doi.org/10.1177/02783649030227002

[5] E.M. Arkin, S.P. Fekete, J.S.B. Mitchell, Approximation algorithms for lawn mowing and milling☆ Comput. Geom. 17 (2000) 25 Comput. Aided Design, 50. https://doi.org/10.1016/S0925-7721(00)00015-8

[6] T. Yang, J.V. Miro, M. Nguyen, Y. Wang, R. Xiong, Template-Free Nonrevisiting Uniform Coverage Path Planning on Curved Surfaces. IEEE ASME Trans. Mechatron. 28 (2023) 1853–1861. https://doi.org/10.1109/TMECH.2023.3275214

[7] M.B. Bieterman, D.R. Sandstrom, A Curvilinear Tool-Path Method for Pocket Machining. J. Manuf. Sci. Eng. 125 (2003) 709–715. https://doi.org/10.1115/1.1596579

[8] M. Held, S. de Lorenzo, On the generation of spiral-like paths within planar shapes. J. Comput. Des. Eng. 5 (2018) 348–357. https://doi.org/https://doi.org/10.1016/j.jcde.2017.11.011

[9] C. Sun, Y. Altintas, Chatter free tool orientations in 5-axis ball-end milling. Int. J. Mach. Tools Manuf. 106 (2016) 89–97. https://doi.org/10.1016/j.ijmachtools.2016.04.007

[10] X. Ban, M. Goswami, W. Zeng, X. Gu, J. Gao, Topology dependent space filling curves for sensor networks and applications, 2013 Proceedings IEEE INFOCOM 2166–2174.

[11] M.M. Nasser, Numerical Computing of Preimage Domains for Bounded Multiply Connected Slit Domains. J. Sci. Comput. 78 (2019) 582–606. https://doi.org/10.1007/s10915-018-0784-9

[12] K. Wu, Y. Lu, Numerical computation of preimage domains for spiral slit regions and simulation of flow around bodies. Math Biosci Eng, 20 (2023) 720–736. https://doi.org/10.3934/mbe.2023033

[13] X. Yin, J. Dai, S.T. Yau, X. Gu, Slit Map: Conformal Parameterization for Multiply Connected Surfaces, in: F. Chen, B. Jüttler (Eds.), Advances in Geometric Modeling and Processing, Springer, Berlin, Heidelberg. 2008, pp. 410–422. https://doi.org/10.1007/978-3-540-79246-8_31

[14] C. Shen, S. Mao, B. Xu, Z. Wang, X. Zhang, S. Yan, H. Ding, Spiral complete coverage path planning based on conformal slit mapping in multi-connected domains. Int. J. Robot. Res. 43 (2024) 2183–2203. https://doi.org/10.1177/02783649241251385

[15] J.J. Chuang, D.C.H. Yang, A laplace-based spiral contouring method for general pocket machining. Int. J. Adv. Manuf. Technol. 34 (2007) 714–723. https://doi.org/10.1007/s00170-006-0648-6

[16] Y. Sun, J. Xu, C. Jin, D. Guo, Smooth tool path generation for 5-axis machining of triangular mesh surface with nonzero genus. Comput. Aided Des. 79 (2016) 60–74. https://doi.org/10.1016/j.cad.2016.06.001

[17] G.P.T. Choi, L.M. Lui, Recent Developments of Surface Parameterization Methods Using Quasi-conformal Geometry, in: K. Chen, C.B. Tai, X.C. Schönlieb, L. Younces (Eds.), Handbook of Mathematical Models and Algorithms in Computer Vision and Imaging, Springer, Cham, 2023, pp. 1483–1523. https://doi.org/10.1007/978-3-030-98661-2_113

[18] C. Yuan, N. Cao, Y. Shi, Complex Surface Fabrication via Developable Surface Approximation: A Survey. IEEE Trans. Vis. Comput. Graph. (2025) 1–20. https://doi.org/10.1109/TVCG.2025.3538782

[19] A. Nojoomi, J. Jeon, K. Yum, 2D material programming for 3D shaping. Nat. Commun. 12 (2021) 603. https://doi.org/10.1038/s41467-021-20934-w

[20] A.V. Shembekar, Y.J. Yoon, A. Kanyuck, S.K. Gupta, Generating Robot Trajectories for Conformal Three-Dimensional Printing Using Nonplanar Layers. J. Comput. Inf. Sci. Eng. 19 (2019) 031011. https://doi.org/10.1115/1.4043013

[21] B. Lévy, S. Petitjean, N. Ray, J. Maillot, Least squares conformal maps for automatic texture atlas generation, in: M.C. Whitton (Ed.), Seminal Graphics Papers: Pushing the Boundaries, Volume 2. Association for Computing Machinery, New York, NY, United States. https://doi.org/10.1145/3596711.3596734

[22] R. Sawhney, K. Crane, Boundary First Flattening. ACM Trans. Graph. 37 (2017) 1–14.

https://doi.org/10.1145/3132705

[23] K. Su, L. Cui, K. Qian, N. Lei, J. Zhang, M. Zhang, X.D. Gu, Area-preserving mesh parameterization for poly-annulus surfaces based on optimal mass transportation. Comput. Aided Geom. Des. 46 (2016) 76–91. https://doi.org/10.1016/j.cagd.2016.05.005

[24] K. Hormann, G. Greiner, MIPS: An Efficient Global Parametrization Method. Curve and Surface Design: Saint-Malo 2000 (2012) 10.

[25] B.H. Kim, B.K. Choi, Machining efficiency comparison direction-parallel tool path with contour-parallel tool path. Comput. Aided Des. 34 (2002) 89–95. https://doi.org/10.1016/S0010-4485(00)00139-1

[26] P. Hu, L. Chen, J. Wang, K. Tang, Boundary-Conformed Tool Path Generation Based on Global Reparametrization. In 2015 14th International Conference on Computer-Aided Design and Computer Graphics (CAD/Graphics), (2015) 165–172.

[27] Z. Liu, X. Li, B. Yi, Generating spiral tool path to machine free-form surface with complex topology based on fusing constraint mapping and enriched Voronoi diagram. Int. J. Adv. Manuf. Technol. 102 (2019) 647–658. https://doi.org/10.1007/s00170-018-3183-3

[28] A.A.M. Yunus, A.H.M. Murid, M.M.S. Nasser, Numerical conformal mapping and its inverse of unbounded multiply connected regions onto logarithmic spiral slit regions and straight slit regions. Proc. R. Soc. A: Math. Phys. Eng. Sci. 470 (2014) 20130514. https://doi.org/10.1098/rspa.2013.0514

[29] G.P.T. Choi, Efficient Conformal Parameterization of Multiply-Connected Surfaces Using Quasi-Conformal Theory. J. Sci. Comput. 87 (2021) 70. https://doi.org/10.1007/s10915-021-01479-y

[30] T. Zhao, Z. Yan, L. Wang, R. Pan, X. Wang, K. Liu, K. Guo, Q. Hu, S. Chen, Hybrid path planning method based on skeleton contour partitioning for robotic additive manufacturing. Robot. Comput. Integr. Manuf. 85 (2024) 102633. https://doi.org/https://doi.org/10.1016/j.rcim.2023.102633

[31] T. Kim, Constant cusp height tool paths as geodesic parallels on an abstract Riemannian manifold.

Comput. Aided Des. 39 (2007) 477–489. https://doi.org/10.1016/j.cad.2007.01.003

[32] Q. Zou, J. Zhang, B. Deng, J. Zhao, Iso-level tool path planning for free-form surfaces. Comput. Aided Des. 53 (2014) 117–125. https://doi.org/10.1016/j.cad.2014.04.006

[33] E. Lee, Contour offset approach to spiral toolpath generation with constant scallop height. Comput. Aided Des. 35 (2003) 511–518. https://doi.org/10.1016/S0010-4485(01)00185-3

[34] Z. Lin, J. Fu, H. Shen, W. Gan, A generic uniform scallop tool path generation method for five-axis machining of freeform surface. Comput. Aided Des. 56 (2014) 120–132. https://doi.org/10.1016/j.cad.2014.06.010

[35] C. Tournier, C. Lartigue, 5-axis Iso-scallop Tool Paths along Parallel Planes. Comput. Aided Des. Appl. 5 (2008) 278–286. https://doi.org/10.3722/cadaps.2008.278-286

[36] S. Pratt, T. Kosmal, C. Williams, Adaptively sampled distance functions: A unifying digital twin representation for advanced manufacturing. Robot. Comput. Integr. Manuf. 92 (2025) 102877. https://doi.org/10.1016/j.rcim.2024.102877

[37] X. Li, H. Yan, J. Sun, Y. Li, G. Zhang, A curved layering algorithm based on voxelization and geodesic distance for robotic GMA additive manufacturing. Virtual Phys. Prototyp. 19 (2024) e2346289. https://doi.org/10.1080/17452759.2024.2346289

[38] R. Kress, A Nyström method for boundary integral equations in domains with corners. Numer. Math. 58 (1990) 145–161. https://doi.org/10.1007/BF01385616

[39] S. Rusinkiewicz, Estimating curvatures and their derivatives on triangle meshes. Proceedings. 2nd International Symposium on 3D Data Processing, Visualization and Transmission, 2004. 3DPVT 2004. 2004, pp. 486–493.

**Figure captions**

**Fig. 1.** Transformation relationships between different mappings: **(a)** The original surface $S$, **(b)** The flattened surface $S^F$ obtained from the surface $S$, **(c)** The iso-scallop surface $S^h$ obtained by offsetting the surface $S$ by a distance $h$ normal to the surface, **(d)** The surface $S^F$ mapped to a disk or annular region $S^S$ using conformal slit mapping, with $O^F$ positioned at different locations.

**Fig. 2.** Tool trajectory generation based on conformal slit mapping and corresponding machining scallop height. (a) and (c) Conformal slit mapping domain $S^S$ and iso-parametric curves on the mapped domain $S^S$. (b) and (d) Tool center trajectories $C_i^T$, tool contact trajectories$C_i$, tool axis direction $\overrightarrow{C_i^A}$, and the corresponding milling scallop height distribution and milling bands $BP_i$.

**Fig. 3.** Spiral bridging of iso-parametric trajectories. (a) Bridging trajectory on $S^R$ that avoids linear slits. (d) Spiral tool trajectory obtained by mapping the bridging trajectory on $S^R$ mapping $\omega_R^{-1}\omega_{SF}^{-1}\omega_I(\cdot, K_c)$, along with the corresponding milling band on $S^h$. (b) and (e) Increasing $D_{3-3}^{Real}$ on $S^R$ to eliminate unswept regions between the milling bands of the corresponding spiral trajectories. (c) and (f) Increasing $D_{4-5}^{Real}$ on $S^R$ to prevent the corresponding spiral trajectories from passing through holes on the surface.

**Fig. 4.** Relationship between $CL_i$, $CL_{i+1}$, $h$, $\nabla T(P_i)$, $K_s$ and $K_c$.

**Fig. 5.** Results of the traversal calculation of $E_{min}^S$ on Surface $S^F$.

**Fig. 6.** Optimization of the origin-mapped point for a complex asymmetric surface.

**Fig. 7.** Comparison of milling trajectory generation using various methods, conformal flattening of low-quality meshes, and parameterization of high-genus surfaces.

**Fig. 8.** Milling trajectories and states obtained using the parameters set in Table 1.

**Fig. 9.** Optimization of the origin-mapped point for a centrally symmetric surface and its effect on trajectory generation. (a) Centrally symmetric surface with symmetry center $O$. (b)

Generated trajectory based on the optimized origin position $\omega_F^{-1}(O_{end-1}^F)$. (c) Convergence of iterative searches to $O_{opt}^F$ from different initial points on the surface.

**Fig. 10.** Experimental platforms: (a) Machining experiment platform, (b) Measurement experiment platform.

**Fig. 11.** Workpieces and scanned scallop height distribution, with red points indicating maximum scallop height sampling locations. (a) and (c) correspond to Workpiece 1, while (b) and (d) correspond to Workpiece 2.

**Fig. 12.** Total acceleration of the Spindle: (a) During machining of Workpiece 1, (b) During machining of Workpiece 2.

**Appendix A**

**Pseudocode A-1.** The pseudocode for the trajectories spacing control.

| | | |
|---|---|---|
| **algorithm:** | Calculating trajectories with appropriate adjacent spacing | **Complexity:** |
| **input:** | The surface $S$ is represented as a triangular mesh, with its corresponding flattened version denoted as $S^F$ and the conformal slit-mapped version as $S^S$. Each of these meshes consists of $N_F$ facets. Let $h$ be the maximum allowable scallop height, $R_c$ the tool radius, and $\varepsilon$ the permissible error in trajectory spacing. To analyze the iso-scallop height surface $S^h$, a total of $N_S$ points are sampled, forming the point set $P^h$. Additionally, the number of discrete points used for iso-parameterization trajectory generation is represented as $N_C$. | |
| **output:** | Tool contact paths $TC$, tool centroid trajectories $TB$, and tool orientation directions $TA$; iso-parametric curve radius $R_k$ on $S^S$; and machining coverage regions $BP$ corresponding to each trajectory. | |
| 1 | Setting $R_{up} = 1$, $R_{down} = 0$ for disk slit mapping, $R_{down} = R_A$ for annular slit mapping; $k = 0$; $TA = TB = TC = BP = R_k = \{\}$ ; Setting $C_{k_old}^T = \omega_{SF}^{-1}\omega_I(C_k^S(R_{down}), R_c)$. | |
| 2 | **while** $P^h \neq \emptyset$ **do** | $\boldsymbol{O(k)}$ |
| 3 | $R_{up_t} = R_{up}$; $R_{down_t} = R_{down}$; $k = k + 1$; $R_{k_t} = \frac{R_{up}+R_{down}}{2}$; $C_k^T = \emptyset$; $P^{h,C_1^T,R_C^+} = \emptyset$. | |

**while** 1 **do** $\boldsymbol{O\left(log\left(\frac{1}{\varepsilon}\right)\right)}$

$C_k^T = \omega_{SF}^{-1}\omega_I\left(C_k^S(R_{k_t}), R_c\right).$ $O(N_C log(N_F))$

Construct a KD-tree for the discrete points of $C_k^T$. $O(N_C log(N_C))$

Calculate $P^{h,C_k^T,R_C^+} = \left\{P_i^h \in P^h \mid \left\|P_i^h - C_k^T\right\|_2 > R_c\right\}$ by KD-tree of $C_k^T$. $\boldsymbol{O(N_S log(N_C))}$

$P^S = \omega_I^{-1}\omega_{FS}(P^h, O^F)$ $O(N_S)$

$P^{S,C_k^T,R_C^+} = \omega_I^{-1}\omega_{FS}\left(P^{h,C_k^T,R_C^+}, O^F\right)$ $O(N_S)$

$P^{P^S,C_k^S} = \left\{P_i^S \in P^{S,C_1^T,R_C^+} \mid \left\|P_i^S - O^S\right\|_2 > R_1^S\right\}$ $O(N_S)$

**if** $P^{P^S,C_k^S} = \emptyset$ **then**

$R_{up_t} = R_{k_t}$

**if** $distance\left(C_{k_old}^T, C_k^T\right) < \varepsilon$ **then**

**break**

**end**

**else**

$R_{down_t} = R_{k_t}$

**end**

$R_{k_t} = \frac{R_{up_t}+R_{down_t}}{2}$; $C_{k_old}^T = C_k^T$.

**end**

$R_{up} = R_{k_t}$; $P^h = P^{h,C_1^T,R_C^+}$; $BP_i = P^h - P^{h,C_k^T,R_C^+}$.

Calculate $\omega_{SF}^{-1}\left(C_k^S(R_k^S)\right)$ and $\overrightarrow{C_k^A}$ by $R_{k_t}$.

$TA\{k\} = \overrightarrow{C_k^A}$; $TB\{k\} = C_k^T$; $TC\{k\} = \omega_{SF}^{-1}\left(C_k^S(R_k^S)\right)$;

$R_k\{k\} = \{R_{k_t}\}$; $BP\{k\} = BP_i$.

23 **end**

24 $R_k\{k+1\} = R_{down}$

25 **return** $TA, TB, TC, R_k, BP$.

---

Complexity analysis of the trajectory spacing control algorithm:

The trajectory spacing regulation algorithm operates through a pair of nested while loops. Within the inner loop (line 4), a binary search mechanism iteratively refines the spacing to determine the subsequent trajectory $C_i^T$, ensuring it remains adjacent to the preceding trajectory $C_{i-1}^T$ while maintaining the prescribed spacing constraints. With an allowable path spacing deviation of $\varepsilon$, the binary search achieves convergence in $O\left(log\left(\frac{1}{\varepsilon}\right)\right)$ time. Each binary search iteration necessitates evaluating the machining band of $C_i^T$, which requires computing the distances between all points in $P^h$ and $C_i^T$. Given that $P^h$ consists of $N_S$ points and $C_i^T$ is discretized into $N_C$ points, constructing a KD-tree for the discrete points of $C_i^T$ enhances computational efficiency, reducing the complexity to $O(N_S log(N_C))$.

The outer while loop (line 2) iterates to generate all trajectories with controlled spacing, where the total number of trajectories is represented by $k$. As the iterations progress, the number of points in $P^h$ gradually diminishes until it reaches zero. However, across all iterations, the average value of $N_S$ stabilizes at roughly one-third of its initial value. Given this, treating $N_S$ as a constant does not impact the complexity assessment. Consequently, the overall computational complexity of the algorithm, derived from the nested loop structure, is expressed as $O\left(kN_S log(N_C) log\left(\frac{1}{\varepsilon}\right)\right)$.

**Pseudocode A-2.** The pseudocode for spiral bridging of iso-parametric trajectories.

---

| | |
|---|---|
| **algorithm:** | Spiral bridging of iso-parametric trajectories |
| **input:** | The triangular mesh representing surface $S$ and its corresponding |

conformal slit-mapped mesh, denoted as $S^S$, serve as the foundation for analyzing trajectory-based machining characteristics. The radii associated with the iso-parametric lines of trajectory $TB$ on $S^S$ are represented by the set $R_k = \{R_1^S, \ldots, R_k^S, R_{k+1}^S\}$ . Similarly, the machining bands that correspond to trajectory $TB$ on the iso-scallop height surface $S^h$ are denoted by the set $BP = \{BP_1, BP_2, \ldots, BP_k\}$.

**output:** The bridged spiral trajectory $L_{best}$.

---

Transform $S^S$ into $S_0^R$ using Equation. 7, then translate $S_0^R$ along the real axis with a period of 2π to obtain $S^R = S_0^R \cup S_1^R \cup S_2^R \cup \ldots$.

Set $L_best = \emptyset$ , and the total length of the current spiral trajectory $LL_best = \infty$.

**for** $real(P_{Start}^1) = 0$ **to** $2\pi$ **with step** $\frac{\pi}{50}$

  $imag(P_{Start}^1) = R_1^S$,$P_{End}^1 = P_{Start}^1 - 2\pi$; Let $L_1^1$ be the straight line connecting $P_{End}^1$ and $P_{Start}^1$.

  **for** $i = 1$ **to** $k$

    **if** $R_i^S > 0.3$

      $D_{i-i+1}^{Real} = \frac{\pi}{10}$; $D_{i+1-i+1}^{Real} = \frac{8\pi}{5}$;

    **else**

      $D_{i-i+1}^{Real} = 2\pi$; $D_{i+1-i+1}^{Real} = 0$;

    **end**

    $D_{i-i+1}^{Real_t} = D_{i-i+1}^{Real}$

    **while** 1 **do**

      $real(P_{End}^{i+1}) = D_{i-i+1}^{Real_t} + real(P_{Start}^i)$; $imag(P_{End}^{i+1}) = R_{i+1}^S$;

      Let $L_{i+1}^i$ be the curve connecting $P_{Start}^i$ and $P_{End}^{i+1}$ through

Equation. 9.

**if** $L_{i+1}^{i}$ does not intersect with the slit on $S^{R}$

**break**;

**end**

$D_{i-i+1}^{Real_t} = D_{i-i+1}^{Real_t} + \frac{\pi}{50}$;

**end**

$D_{i+1-i+1}^{Real_t} = D_{i+1-i+1}^{Real}$;

**while** 1 **do**

$P_{Start}^{i+1} = P_{End}^{i+1} + D_{i+1-i+1}^{Real_t}$; Let $L_{i+1}^{i+1}$ be the straight line connecting $P_{End}^{i+1}$ and $P_{Start}^{i+1}$.

$real(P_{End}^{i+2}) = real(P_{Start}^{i+1}) + D_{i-i+1}^{Real}$; $imag(P_{End}^{i+2}) = R_{i+2}^{S}$; Let $L_{i+2}^{i+1}$ be the curve connecting $P_{Start}^{i+1}$ and $P_{End}^{i+2}$ through Equation. 9.

Calculate whether the region swept by the ball end mill along the trajectory $\omega_{SF}^{-1}\omega_{I}\left(\omega_{R}^{-1}\left(L_{i}^{i} \cup L_{i+1}^{i} \cup L_{i+1}^{i+1} \cup L_{i+2}^{i+1}\right), R_{c}\right)$ completely sweeps the milling band $BP_{i+1}$.

**if** $BP_{i+1}$ is completely swept

**break**;

**end**

$D_{i+1-i+1}^{Real_t} = D_{i+1-i+1}^{Real_t} + \frac{\pi}{50}$;

**end**

**end**

Set the total length of $\omega_{SF}^{-1}\omega_{I}\left(\omega_{R}^{-1}\left(L_{1}^{1} \cup L_{2}^{1} \cup L_{2}^{2} \cup L_{3}^{2} \cup ... \cup L_{k}^{k} \cup\right.\right.$

```
    L^k_{k+1}), R_c) as LL_best_t.
    if LL_best_t < LL_best
        L_best = {L^1_1 ∪ L^1_2 ∪ L^2_2 ∪ L^2_3 ∪...∪ L^{k-1}_{k-1} ∪ L^{k-1}_k} ; LL_best = LL_best_t.
    end
end
return L_best
```

**Appendix B**

Let $S$ be a two-dimensional surface, and consider the objective of minimizing the functional:

$$\mathcal{J}(f) = \int_S \left(f(x) + \frac{1}{f(x)}\right) dS(x) \tag{B-1}$$

where $dS$ denotes the surface element on $S$, and $f$ is strictly positive over $S$. Additionally, the function $f$ must satisfy the constraint that its average value over the surface is given by $Avg$, expressed as:

$$\frac{1}{A_S} \int_S \big(f(x)\big)\, dS(x) = Avg \tag{B-2}$$

where $A_S$ represents the area of the surface. To determine the function $f$ that minimizes the integral in Equation (B-1) while satisfying the constraint in Equation (B-2), the method of Lagrange multipliers is employed to construct the Lagrangian function:

$$\mathcal{L}(f) = \int_S \left(f(x) + \frac{1}{f(x)}\right) dS(x) + \lambda \left(\frac{1}{A_S} \int_S \big(f(x)\big)\, dS(x) - Avg\right) \tag{B-3}$$

where $\lambda$ represents the Lagrange multiplier. Taking the functional derivative of $\mathcal{L}(f)$ with respect to $f(x)$ gives:

$$\frac{\delta}{\delta f(x)} \mathcal{L}(f) = 1 - \frac{1}{f(x)^2} + \frac{\lambda}{A_S} \tag{B-4}$$

which leads to the Euler-Lagrange equation:

$$1 - \frac{1}{f(x)^2} + \frac{\lambda}{A_S} = 0 \tag{B-5}$$

Solving for $f(x)$, the expression simplifies to:

$$f(x) = \frac{1}{\sqrt{1 + \frac{\lambda}{A_S}}} \tag{B-6}$$

Therefore, the optimal function satisfying both the minimization objective and the constraint is the constant function $f(x) = Avg$.

## Appendix C

To compute the discrete form of $E^S$ when the function $f$ is known, consider a triangular face $F_i$ within the mesh $S$, defined as:

$$F_i = \{V_i^1, V_i^2, V_i^3\}, \text{ where } i = 1,2,\dots,N_F \tag{C-1}$$

where $N_F$ represents the total number of triangular faces in the mesh. The three vertices of $F_i$ are $\{V_i^1, V_i^2, V_i^3\}$. When mapped via $\omega_{FS}$, this triangular face corresponds to another $S^S$, denoted as:

$$FS_i = \{VS_i^1, VS_i^2, VS_i^3\} \tag{C-2}$$

As established in Section 2.3.1, knowing $f$ allows the determination of the corresponding $T$ at each node of the mesh $S$. The average gradient over the triangular face $F_i$, represented as $\nabla T(F_i)$, can then be expressed as follows:

$$\nabla T(F_i) = \left( (T(V_i^2) - T(V_i^1)) \frac{\left(\overrightarrow{V_i^3 V_i^1}\right)^{\perp}}{2A_{F_i}} \right) + \left( (T(V_i^3) - T(V_i^1)) \frac{\left(\overrightarrow{V_i^2 V_i^1}\right)^{\perp}}{2A_{F_i}} \right) \tag{C-3}$$

The area of the triangular face $F_i$, denoted as $A_{F_i}$, serves as a key parameter in curvature calculations. When considering two mutually perpendicular unit vectors $\left\{\overrightarrow{u_{F_i}}, \overrightarrow{u_{F_i}}^{\perp}\right\}$ that define a local coordinate system on $F_i$, the curvature along the gradient direction $\nabla T(F_i)$ can be derived using the second fundamental form parameters $\left\{E_{F_i}, F_{F_i}, G_{F_i}\right\}$ [39]. This formulation allows for the computation of the average curvature $K_s\left(\nabla T(F_i)\right)$ by integrating these coefficients with respect to the directional gradient.

$$K_s\left(\nabla T(F_i)\right) = E_{F_i} \frac{\left\|\nabla T(F_i)\overrightarrow{u_{F_i}}\right\|_2^2}{\|\nabla T(F_i)\|_2^2} + 2F_{F_i} \frac{\left\|\nabla T(F_i)\overrightarrow{u_{F_i}}\right\|_2 \left\|\nabla T(F_i)\overrightarrow{u_{F_i}}^{\perp}\right\|_2}{\|\nabla T(F_i)\|_2^2} + G_{F_i} \frac{\left\|\nabla T(F_i)\overrightarrow{u_{F_i}}^{\perp}\right\|_2^2}{\|\nabla T(F_i)\|_2^2} \tag{C-4}$$

Therefore, the discrete form $E^S$ can be expressed in the form below:

$$E^S = \sum_{i=1}^{N_F} A_{F_i} \left( \frac{K_s\left(\nabla T(F_i)\right) + K_c}{8\|\nabla T(F_i)\|_2^2} + \frac{8\|\nabla T(F_i)\|_2^2}{K_s\left(\nabla T(F_i)\right) + K_c} \right) \tag{C-5}$$

The following functional problem was then solved to minimize $E^S$ under the given $O^F$:

$$\begin{cases} f^* = argmin_{f\in\{f|f(R_{min})=0, f'(x)>0 \text{ for } x\in[R_{min},1]\}} E^S(f, O^F) \\ E^S_{min}(O^F) = E^S(f^*(O^F), O^F) \end{cases} \tag{C-6}$$

Denote the $p + 1$ nodes on $f$ as:

$$P_f = \begin{bmatrix} P_0 & P_1 & \cdots & P_p \\ f(P_0) & f(P_1) & \cdots & f(P_p) \end{bmatrix} \tag{C-7}$$

where $P_i = R_{min} + j\frac{(1-R_{min})}{p}$ for $j = 0,1,\ldots,p$. The function $f$ can be approximately represented by linear interpolation of $P_f$ as:

$$f(x) \approx \sum_{i=0}^{p-1}\left[f(P_i) + \frac{x-P_i}{P_{i+1}-P_i}\big(f(P_{i+1}) - f(P_i)\big)\right]H_{[P_i,P_{i+1}]}(x),\ x \in [R_{min}, 1] \tag{C-8}$$

where:

$$H_{[P_i,P_{i+1}]}(x) = \begin{cases} 1\ for\ x \in [P_i, P_{i+1}] \\ 0\ for\ x \notin [P_i, P_{i+1}] \end{cases} \tag{C-9}$$

If $f(P_0) = 0$ and $f(P_j) < f(P_{j+1})$, then the interpolation function adheres to the conditions $f(R_{min}) = 0$ and $f'(x) > 0$ over the interval $[R_{min}, 1]$. Introducing a perturbation $\delta f(P_j)$ at $f(P_j)$ affects the function solely within the range $I_j = [I_j^{min}, I_j^{max}] \subset [R_{min}, 1]$ under linear interpolation. Here, $j = 1,2,3,\ldots p$, $I_j = [P_{j-1}, P_{j+1}]$ for $0 < j < p$, and $I_j = [P_{p-1}, P_p]$ when $j = p$.

Define an annular region on $S^S$, covering the radius interval $[I_j^{min}, I_j^{max}]$, as $An_j = \{Z \subset S^S | I_j^{min} \le \|Z\|_2 \le I_j^{max}\}$. Let the set of triangular faces contained within this region be represented as $\{FS^j\} = \{FS_j^k \in \{FS_1, FS_2, \ldots, FS_{N_F}\} | An_j \cap FS_j^k \ne \emptyset\}$, where $N_F^j$ is the number of triangular elements in $\{FS^j\}$. The corresponding triangular facets in $S^S$ are denoted by $F_j^k$, each with an associated area $A_j^k$. The resulting variation in the discrete form $E^S$ due to the perturbation $\delta f(P_i)$ is:

$$\delta E^S\left(\delta f(P_j)\right) = \sum_{k=1}^{N_F^j} A_j^k \left[\frac{8\left\|\nabla(T+\delta T)\left(F_j^k\right)\right\|_2}{K_s\left(\nabla(T+\delta T)\left(F_j^k\right)\right)+K_c} + \frac{K_s\left(\nabla(T+\delta T)\left(F_j^k\right)\right)+K_c}{8\left\|\nabla(T+\delta T)\left(F_j^k\right)\right\|_2} - \frac{8\left\|\nabla(T)\left(F_j^k\right)\right\|_2}{K_s\left(\nabla(T)\left(F_j^k\right)\right)+K_c} - \right.$$

$$\left.\frac{K_s\left(\nabla(T)\left(F_j^k\right)\right)+K_c}{8\left\|\nabla(T)\left(F_j^k\right)\right\|_2}\right] \quad \text{(C-10)}$$

When $T + \delta T$ represents the interpolated temperature at the mesh nodes after introducing the perturbation $\delta f(P_j)$ to $f$, the computational complexity of Equation C-10 is given by $O(N_F^j)$. For $j = 1, \ldots, p - 1$, consider the following three types of perturbations applied to $f(P_j)$:

$$\{\delta f(P_j^1), \delta f(P_j^2), \delta f(P_j^3)\} = \left\{\frac{f(P_{j-1}) - f(P_j)}{2D_f(Idx)}, 0, \frac{f(P_{j+1}) - f(P_j)}{2D_f(Idx)}\right\} \quad \text{(C-11)}$$

where $Idx$ represents the iteration count of $f$, while the exponential decay function is defined as $D_f(Idx) = (1.01)^{Idx}$. The variations in energy, denoted as $\left\{\delta E^S\left(\delta f(P_j^1)\right), 0, \delta E^S\left(\delta f(P_j^3)\right)\right\}$, resulting from the three perturbations, can be determined using Equation C-11. To approximate these changes, a quadratic function $Qf_j(x) = Ax^2 + Bx + C$ is fitted based on three points, $Q_j$, defined as follows:

$$Q_j = \begin{bmatrix} \frac{f(P_{j-1}) - f(P_j)}{2D_f(Idx)} & 0 & \frac{f(P_{j+1}) - f(P_j)}{2D_f(Idx)} \\ \delta E^S\left(\delta f(P_j^1)\right) & 0 & \delta E^S\left(\delta f(P_j^3)\right) \end{bmatrix} \quad \text{(C-12)}$$

The minimum value of the quadratic function $Qf_j(x)$, denoted as $Qf_j^{min}$, along with the corresponding variable $x_j^{min}$ within the interval $QI = [\delta f(P_j^1), \delta f(P_j^3)]$, can be expressed as follows:

$$\begin{cases} Qf_j^{min} = min_{x \in QI} Qf_j(x) \\ x_j^{min} = argmin_{x \in QI} Qf_j(x) \end{cases} \quad \text{(C-13)}$$

Define $x_j^{min}$ as the optimal perturbation magnitude added to $f(P_j)$ at each iteration, ensuring that $\delta f(P_j) = x_j^{min}$, for $j = 0, \ldots, p - 1$ during each iteration. For the case where $j = p$, introduce seven distinct perturbations to $f(P_p)$:

$$\begin{cases} \delta f(P_p^k) = \left(f(P_{p-1}) - f(P_p)\right)(1 - 0.9^k), k = 1,2,3 \\ \delta f(P_p^k) = (1.1^{k-4} - 1) f(P_0), k = 4,5,6,7 \end{cases} \quad \text{(C-14)}$$

The optimal perturbation magnitude applied to $f(P_p)$ in each iteration is given by $\delta f(P_p) = \delta f(P_p^{k_{min}})$, where $k_{min} = argmin_{k\in 1,2,\ldots,7}\delta E^S\left(\delta f(P_p^k)\right)$.

At this stage, the optimal perturbation value $\delta P_f$ for each iteration has been established at the interpolation points $P_f$ of the linear approximation function $f$.

$$\delta P_f = \begin{bmatrix} 0 & 0 & \cdots & 0 & 0 \\ 0 & \delta f(P_1) & \cdots & \delta f(P_{p-1}) & \delta f(P_p) \end{bmatrix} \tag{C-15}$$

Let $P_f = P_f + \delta P_f$, and repeat the update until the following perturbation energy converges:

$$\sum_{j=1}^{p} \delta E^S\left(\delta f(P_j)\right) < E_\varepsilon^S \tag{C-16}$$

The allowable error for energy convergence is denoted as $E_\varepsilon^S$. The function $f^*$ is approximated through the linear interpolation of $P_f$ after achieving perturbation energy convergence. The computational complexity for each iteration of $\delta P_f$ is expressed as:

$$O\left(7N_F^p + 3\sum_{j=1}^{p-1} N_F^j\right) \tag{C-17}$$

Based on the coverage of $An_j$, it follows that the summation $\sum_{j=1}^{p-1} N_F^j \approx 2N_F$ and $N_F^p << N_F$. Consequently, the time complexity in Equation (C-17) can be estimated as:

$$O\left(7N_F^p + 3\sum_{j=1}^{p-1} N_F^j\right) = O(N_F) \tag{C-18}$$

Figure C-1 presents the convergence behavior of $E^S$ and the iterative adaptation of $f$ under varying initial conditions. The analysis of Figures C-1(a) and C-1(b) reveals that $E^S$ undergoes exponential convergence irrespective of the starting values of $f$. Similarly, Figures C-1(c) and C-1(d) demonstrate that the iteration results for ff ultimately align, regardless of its initial state. These observations highlight the robustness of the algorithm and its rapid convergence properties. When the permissible convergence error for $E^S$ is given by $E_\varepsilon^S$, the number of iterations required for $E^S$ to reach convergence is approximately on the order of $log\left(\frac{1}{E_\varepsilon^S}\right)$. Consequently, the overall computational complexity of the $E_{min}^S$ algorithm is expressed as

$O\left(N_F log\left(\frac{1}{E_\varepsilon^S}\right)\right)$.

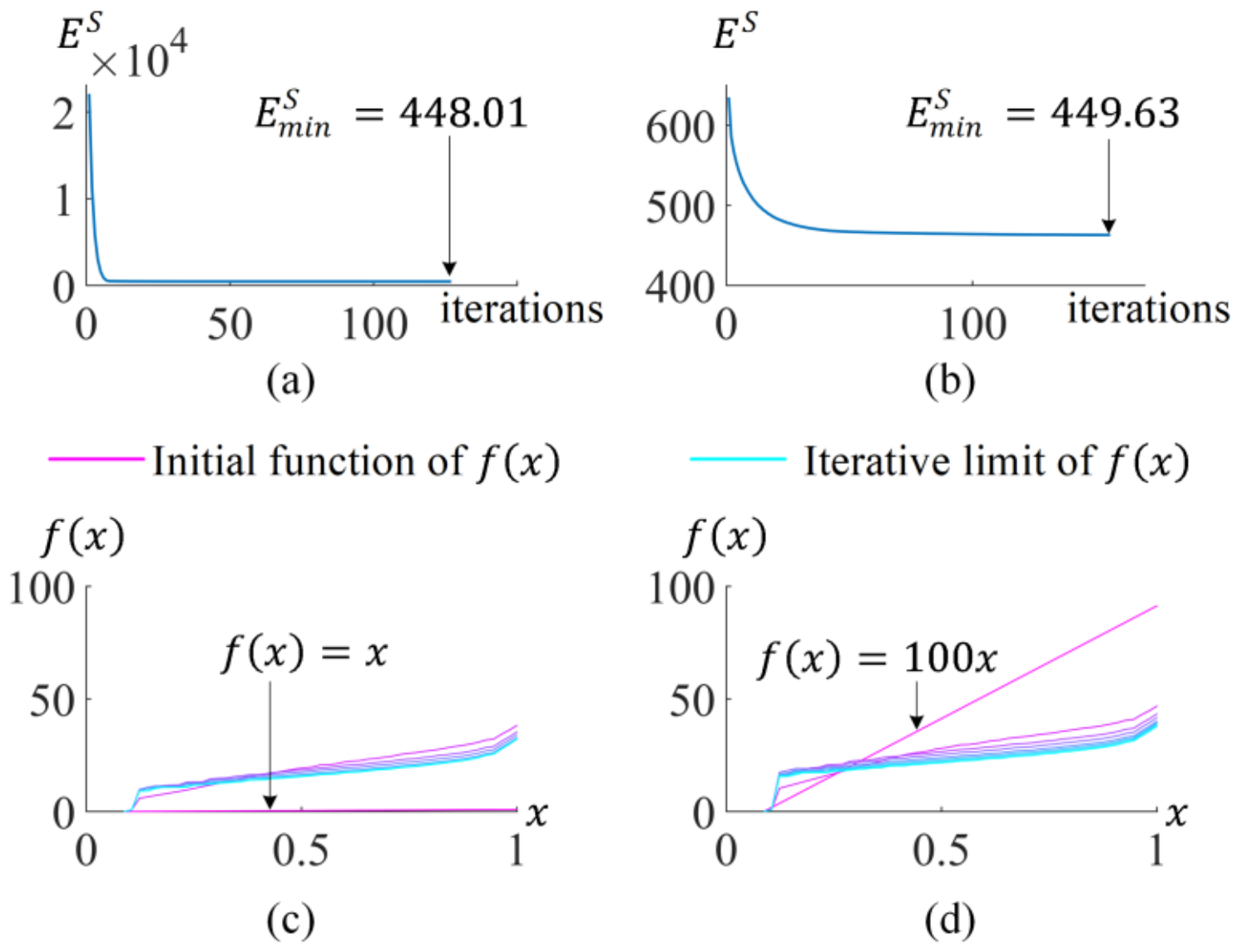

**Fig. C-1.** (a) Iteration process of $E^S$ when the initial value of $f$ is $f(x) = x$. (b) Iteration process of $E^S$ when the initial value of $f$ is $f(x) = 100x$. (c) Iteration process of $f$ itself when the initial value of $f$ is $f(x) = x$. (d) Iteration process of $f$ itself when the initial value of $f$ is $f(x) = 100x$.

**Pseudocode. C-1.** The pseudocode for calculating $E^S_{min}(O^F)$.

| | | |
|---|---|---|
| **algorithm:** | Calculate $E^S_{min}(O^F)$ | **Complexity:** |
| **input:** | The triangular mesh of the surface $S$, the flattened mesh of $S$ denote as $S^F$, $O^F \in S^F(\Gamma_0^F) - \Gamma^F$. | |
| **output:** | $E^S_{min}(O^F)$. | |
| 1 | Calculate $\omega_S: S^F \rightarrow (S^S;\ O^F)$. | |
| 2 | The initial function is $f(x) = x$, where $x \in [0, 1]$ for $S^S$ as a disk, and $x \in [R_A, 1]$ for $S^S$ as an annular region, with $R_A$ representing the inner radius of the annulus. $\delta E^S = \infty$. | |
| 3 | The function $f$ is discretized into several points as a linear interpolation of $P_f$, where: $P_f = \begin{bmatrix} P_0 & P_1 & \cdots & P_p \\ f(P_0) & f(P_1) & \cdots & f(P_p) \end{bmatrix}$. | |
| 4 | **while** $\sum_{j=1}^{p} \left\| \delta E^S\left(\delta f(P_j)\right) \right\| > E^S_\varepsilon$ **do** | $\boldsymbol{O\left(log\left(\frac{1}{E^S_\varepsilon}\right)\right)}$ |
| 5 | **for** $j = 1$ **to** $p - 1$ | $\boldsymbol{O(p-1)}$ |
| 6 | From Equation. C-11, three perturbations $\{\delta f(P_j^1), \delta f(P_j^2), \delta f(P_j^3)\}$ are applied to $f(P_j)$. The corresponding perturbation energies $\left\{\delta E^S\left(\delta f(P_j^1)\right), 0, \delta E^S\left(\delta f(P_j^3)\right)\right\}$ are computed using Equation. C-10. | $\boldsymbol{O\left(\frac{3N_F}{p}\right)}$ |
| 7 | By fitting a quadratic function, the optimal perturbation $\delta f(P_j)$ that minimizes $\delta E^S$ is | |

determined, along with its corresponding perturbation energy $\delta E^S\left(\delta f(P_j)\right)$.

**end**

Apply seven perturbations to $f(P_p)$in Equation. C-14. Using Equation. C-10, determine the index $j$ that minimizes $\delta E^S\left(\delta f(P_p^j)\right)$, then set $f(P_p) = \delta f(P_p^j)$ and $\delta E^S\left(\delta f(P_p)\right) = \delta E^S\left(\delta f(P_p^j)\right)$. $\boldsymbol{O}\left(\frac{\boldsymbol{7N_F}}{\boldsymbol{p}}\right)$

Update $f(P_j) = f(P_j) + \delta f(P_j)$ , where $j = 1,2,\ldots,p$.

**end**

**return** $E_{min}^S(O^F) = E^S$

**Pseudocode. C-2.** The pseudocode for searching the optimal position of $O^F$.

**algorithm:** Search the optimal position of $O^F$.

**input:** The triangular mesh of the surface $S$, the flattened mesh of $S$ denote as $S^F$, $O^F \in S^F(\Gamma_0^F) - \Gamma^F$.

**output:** An optimized point $O_{opt}^F \in S^F(\Gamma_0^F) - \Gamma^F$ is mapped to origin $O^S$ by conformal slit mapping.

---

Arbitrarily generate an initial point $O^F \in S^F(\Gamma_0^F) - \Gamma^F$, set $E_{min_old}^S = 0$, and initialize $E_{min_new}^S = -\infty$.

**while** $E_{min_old}^S - E_{min_new}^S > E_\varepsilon^S$ **do**

  $E_{min_old}^S = E_{min}^S(O^F)$.

  **if** $O^F \in S^F(\Gamma_i^F)$, where $i = 1,2,\ldots,m$.

    Compute the point $O_{i-off}^F$ on the offset curve $\Gamma_{i-off}^F$ of $\Gamma_i^F$ that minimizes $E^S$.

    $O^F = O_{i-off}^F$

  **end**

  Arbitrarily assign a unit vector $\vec{u}$ on $S^F$ and its orthogonal unit vector $(\vec{u})^\perp$, then compute $\nabla E_{min}^S(O^F)$ using Equation. 20.

  $O^F = O^F + \frac{\lambda}{(1.01)^{Idx}} \nabla E_{min}^S(O^F)$; $E_{min_new}^S = E_{min}^S(O^F)$.

**end**

**return** $O_{opt}^F = O^F$